\documentclass[aps,prev,twocolumn,preprintnumbers,floatfix,nofootinbib]{revtex4-1}
\pdfoutput=1
\usepackage{graphicx}
\usepackage{epstopdf}
\usepackage{mathrsfs}
\usepackage{amssymb}
\usepackage{verbatim}
\usepackage{color}
\usepackage{multirow}
\usepackage{orcidlink}

\usepackage{amsmath}

\newcommand{\beq}{\begin{equation}}
\newcommand{\eeq}{\end{equation}}
\newcommand{\ga}{\lower.7ex\hbox{$\;\stackrel{\textstyle>}{\sim}\;$}}
\newcommand{\la}{\lower.7ex\hbox{$\;\stackrel{\textstyle<}{\sim}\;$}}
\newcommand{\trh}{T_{\rm RH}}

\definecolor{c1}{rgb}{0., 0.26, 0.9}
\newcounter{qnumber}

\usepackage{hyperref}
\usepackage{amsmath,bm}
\usepackage{physics}

\hypersetup{
    colorlinks = true,
    citecolor = {blue},
    linkcolor = {blue},
    urlcolor = {blue},
}

\begin{document}

\def\jcap{\ref@jnl{J. Cosmology Astropart. Phys.}}

\vspace{0.5cm}
\title{Constraints on Attractor Models of Inflation and Reheating from \\
{\textit{Planck}}, BICEP/Keck, ACT~DR6, and SPT-3G Data}

\author{John Ellis \orcidlink{0000-0002-7399-0813}}
\email{John.Ellis@cern.ch} 
\affiliation{Theoretical Particle Physics and Cosmology Group, Department of
  Physics, King's~College~London, London WC2R 2LS, UK; \\Theoretical Physics Department, CERN, CH-1211 Geneva 23,
  Switzerland}

\author{Marcos A. G. Garcia \orcidlink{0000-0003-3496-3027}}
\email{marcos.garcia@fisica.unam.mx}
\affiliation{Departamento de F\'isica Te\'orica, Instituto de F\'isica, Universidad Nacional Aut\'onoma de M\'exico, Ciudad de M\'exico C.P. 04510, Mexico} 

\author{Keith~A.~Olive \orcidlink{0000-0001-7201-5998}}
\email{olive@umn.edu}
\affiliation{William I. Fine Theoretical Physics Institute, School of
 Physics and Astronomy, University of Minnesota, Minneapolis, MN 55455,
 USA} 

\author{Sarunas Verner \orcidlink{0000-0003-4870-0826}}
\email{verner@uchicago.edu}
\affiliation{Kavli Institute for Cosmological Physics, \\
University of Chicago, 5640 South Ellis Ave., Chicago, IL 60637, USA}
\vspace{0.5cm}

\date{\today}

\begin{abstract}
We analyze the latest cosmic microwave background (CMB) constraints on the scalar spectral index $n_s$ and tensor-to-scalar ratio $r$ from \textit{Planck} 2018, BICEP/Keck 2018, the Atacama Cosmology Telescope Data Release 6 (ACT DR6), and the South Pole Telescope (SPT-3G) data, focusing on their implications for attractor models of inflation. We compare systematically observational bounds with theoretical predictions for both E-model ($\alpha$-Starobinsky) and T-model potentials. 
The observational constraints accommodate E-models with $\alpha \lesssim 25$, with the canonical Starobinsky model ($\alpha = 1$) predicting $n_s = 0.958-0.963$ for reheating temperatures between $100 - 10^{10}$~GeV, in good agreement with {\it Planck} 2018 data and within the 95\% CL region determined by the {\it Planck}-ACT-SPT combination, but below the 95\% confidence region of the {\it Planck}-ACT-DESI combination. Higher reheating temperatures from near-instantaneous reheating improve the compatibility.  T-models predict slightly lower $n_s$ values (0.956-0.961), in some tension with {\it Planck} 2018 data, and we find an upper limit of $\alpha \lesssim 11$ in these models.  We extend our analysis to generalized $\alpha$-attractors with monomial potentials $V(\varphi) \propto \varphi^k$ near the minimum, demonstrating that models with $k \geq 6$ naturally predict  $n_s \simeq 0.965 - 0.968$ for typical number of $e$-folds, in better agreement with the ACT DR6 data. 
We also consider deformed E- and T-models, which allow significantly higher values of $n_s$ for low values of $\alpha \simeq 1$.

\begin{center}
{\tt UMN-TH-4512/25, FTPI-MINN-25/14, KCL-PH-TH/2025-42, CERN-TH-2025-199} 
\end{center}

\end{abstract}

\maketitle

\section{Introduction}
\label{sec:intro}
Inflation, a period of accelerated expansion in the early Universe, provides a compelling explanation for the origin of primordial density fluctuations and the near-flatness and homogeneity of the observable Universe~\cite{reviews}. These two predictions became testable when observations of the total mass-energy density relative to the critical density corresponding to a flat Universe, $\Omega_{\rm tot}$, 
and the spectral tilt, $n_s$, of the scalar anisotropy spectrum, were determined first by WMAP \cite{wmap} and then with significantly higher precision by \textit{Planck} \cite{Planck}. While most models of inflation predict that $\Omega_{\rm tot}$ should be indistinguishable from unity, measurements of 
the spectral tilt and upper limits on the ratio of tensor and scalar perturbation amplitudes, $r$, have become important discriminators between models~\cite{Planck,BICEP2021,Tristram:2021tvh}.
Many models of inflation could be excluded by their combination. 

Until recently, the CMB data appeared to favor the original inflationary model of Starobinsky~\cite{Staro} as being largely consistent with the determination of $n_s$ and the upper limit on $r$. The Starobinsky model, initially formulated as a means to avoid the initial Big Bang singularity, can be expressed as a modification of the Einstein-Hilbert action for gravity, namely
\beq
{\cal S} \; = \;  \frac{M_P^2}{2}\int d^4x \sqrt{-g} \left(-R + \frac{R^2}{6M^2} \right) \, ,
\label{SStaro}
\eeq
where $R$ is the Ricci scalar, $M$ is the inflationary mass scale, and $M_P = (8\pi G)^{-1/2} \simeq 2.435 \times 10^{18}$ GeV is the reduced Planck mass.

This theory may be rewritten in the Einstein frame as Einstein-Hilbert gravity with a supplementary canonical scalar field, the inflaton, $\varphi$, (sometimes referred to as the scalaron in this context) \cite{WhittStelle,Barrow:1988xh,Kalara:1990ar} with a scalar potential given by
\beq
V \; = \; \frac34 M^2 M_P^2 \left(1 - e^{-\sqrt{\frac23} \frac{\varphi}{M_P}} \right)^2 \, . 
\label{starpot}
\eeq
The inflationary mass scale $M$ (also the inflaton mass for this potential) is determined by the amplitude of the scalar fluctuation spectrum,
\beq
A_s \; = \; \frac{3 M^2}{8\pi^2 M_P^2} \sinh^4 \left(\frac{\varphi_*}{\sqrt{6}M_P} \right) \, , 
\label{AsStaro}
\eeq
where $\varphi_* \simeq 5.35 M_P$ is the value of the inflaton field prior to the final $\sim 55$ $e$-folds of inflation (the determination of $\varphi_*$ will be discussed in greater detail below) and $A_s \simeq 2.1 \times 10^{-9}$ \cite{Planck}, in which case $M = 1.25 \times 10^{-5} M_P \simeq 3 \times 10^{13}$~GeV. The inflationary slow-roll parameters determined from $\varphi_*$ (also discussed below) fix the scalar tilt, $n_s = 0.965$ and a tensor-to-scalar ratio $r = 0.0035$.

The \textit{Planck} satellite experiment~\cite{Planck} in combination with gravitational lensing measured the scalar tilt as
\beq
n_s \; = \; 0.9649 \pm 0.0042 \; (68\%~{\rm CL}) \, ,
\label{nsexp}
\eeq 
and \textit{Planck} in combination with observations by BICEP/Keck~\cite{BICEP2021} 
provided an upper limit on $r$:
\beq
r \; < \; 0.036 \, ,
\label{rlim}
\eeq
both in excellent agreement with the predictions of the Starobinsky model.~\footnote{We note that a more stringent bound of $r < 0.032$ was obtained in \cite{Tristram:2021tvh}.}

Recently, two ground-based CMB experiments have released new results that impact the value  of $n_s$.  Both the Atacama Cosmology Telescope (ACT) \cite{ACT:2025fju,ACT:2025tim} and the South Pole Telescope (SPT) \cite{SPT-3G:2025bzu} have released high resolution data at small angular scales (large multipoles). These alone are neither very different nor more accurate  than the \textit{Planck} result for $n_s$.
However, when combined with \textit{Planck} results that include large angular scales, the derived values of $n_s$ shift (particularly in the combination of \textit{Planck} and ACT data when
DESI DR1 data~\cite{DESI:2024uvr, DESI:2024mwx} and DR2 data~\cite{DESI:2025zpo,DESI:2025zgx}) with smaller uncertainties. 
In the following we will discuss the implications of the BICEP/Keck 2018 results which incorporate \textit{Planck} and WMAP \cite{BICEP2021} with the ACT and SPT results. 

Among the diverse landscape of inflationary theories, no-scale supergravity \cite{no-scale,LN} has emerged as a particularly compelling framework, capable of mimicking the predictions for $n_s$ and $r$ of the Starobinsky model~\cite{ENO6}, and also able to accommodate generalizations such as attractor models~\cite{ENO7,Kallosh:2013hoa,T-model,KLR}. These models naturally produce a flat potential plateau at large field values that generates inflationary dynamics consistent with {\it Planck} measurements.

Two specific types of attractor potentials have been widely studied~\cite{ENO7,KLR,T-model,e-m,Carrasco:2015rva,ENOV3}: $\alpha$-Starobinsky models, which are also referred to as E-models in the literature~\cite{e-m}:
\begin{align}
V & \; = \; \frac{3}{4} \lambda  M_{P}^{4}\left(1-e^{-\sqrt{\frac{2}{3 \alpha}} \frac{\varphi}{M_{P}}}\right)^{2} \, ,
\label{eq:emodel}
\end{align}
and T-Models~\cite{Carrasco:2015rva}:
\begin{align}
\label{eq:tmodel}
V & \; = \; \frac{3}{4}  \lambda M_P^4 \tanh ^{2} \left(\frac{\varphi}{\sqrt{6 \alpha}M_P } \right) \,, 
\end{align}
where $\lambda$ is determined by the CMB normalization and corresponds to $M^2/M_P^2$ in the Starobinsky model, which is a special case of Eq.~(\ref{eq:emodel}) with $\alpha = 1$.~\footnote{We adopt a normalization convention in which the normalization scales of both inflaton potentials coincide, as detailed in Eq.~(\ref{eq:infnorm}). This choice leaves the predicted spectral index $n_s$ and tensor-to-scalar ratio $r$ unchanged.} Generalized versions of these models are discussed in Section \ref{sec:infact}.

In response to the more recent CMB results, 
a wave of studies has re-examined the wider inflationary landscape. 
For example, Ref.~\cite{Kallosh:2025rni} demonstrated that a simple generalization of chaotic inflation with nonminimal coupling to gravity provides a good match to ACT results with $r \simeq 10^{-2}$, while comprehensive analyses of slow-roll models reveal that the combined ACT DR6, \textit{Planck}, DESI, and BICEP/Keck datasets place increased pressure on conventional inflationary scenarios~\cite{Aoki:2025wld, Berera:2025vsu, Brahma:2025dio, Gialamas:2025kef, accid, Salvio:2025izr, Dioguardi:2025mpp, Gao:2025onc, He:2025bli, Drees:2025ngb, Maity:2025czp, Yin:2025rrs, Byrnes:2025kit, Biswas:2025adi, Haque:2025uga, Frolovsky:2025iao, Heidarian:2025drk, Choudhury:2025vso, Dioguardi:2025vci}. The ACT data exclude standard Starobinsky and Higgs inflation models at the 2$\sigma$ level, motivating investigations of reheating effects~\cite{Haque:2025uri, Zharov:2025evb, Liu:2025qca, Mondal:2025kur, German:2025ide}, curvature corrections to Starobinsky inflation that can reconcile predictions with ACT observations~\cite{Kim:2025dyi,Gialamas:2025ofz, Haque:2025uis, Yogesh:2025wak, Addazi:2025qra, Ahmed:2025rrg, Modak:2025bjv,Cheong:2025vmz}, and polynomial potential inflation models that remain viable~\cite{Yi:2025dms, Kallosh:2025ijd, Peng:2025bws}. Studies of some attractor models show that radiative corrections may shift predictions into the ACT-favored parameter space~\cite{Pallis:2025nrv, Chakraborty:2025oyj, Odintsov:2025wai, Wolf:2025ecy, Han:2025cwk, Pallis:2025gii} and smooth hybrid inflation models that naturally predict spectral indices around 0.97~\cite{Okada:2025lpl}, collectively demonstrating how precision cosmological measurements continue to refine theoretical understanding of the early universe and challenge established paradigms in inflationary cosmology. 

Overall, the consensus emerging from recent works is that \emph{modified} plateau potentials, whether via higher-order exponents, non-minimal couplings, or modest multi-field extensions, remain fully compatible with the ACT DR6 + \textit{Planck} + BK18 data, while simple chaotic monomials and the original $R^2$ model are increasingly challenged. The sensitivity of the planned LiteBIRD~\cite{Hazumi:2019lys} mission to $r \sim \mathcal{O}(10^{-3})$ will provide a decisive test for this refined class of inflationary scenarios.

In this paper, we explore the impact of the latest BICEP/Keck/WMAP/\textit{Planck}/ACT/SPT constraints in the $(n_s, r)$ plane on the E-model and T-model inflationary attractor models. We find that the increase in $n_s$ from the {\it Planck}/ACT/DESI DR1 (P-ACT-LB) combination creates significant tension with standard $\alpha$-attractor predictions, whereas the {\it Planck}/ACT/SPT (CMB-SPA) results remain consistent with the conventional expectations.
We examine systematically the allowed values of the $\alpha$ parameter, 
paying particular attention to the reheating temperature that, together with the reheating equation of state $w_{\rm reh}$, fixes the $e$-folding number $N_*$. The allowed range of $N_*$ is determined by the minimum reheating temperature, which we take to be the electroweak scale, $\sim 100$~GeV, and the maximum reheating temperature, which we take as $\sim 10^{10}$~GeV, motivated by supergravity models with a 100 GeV gravitino and supersymmetric dark matter~\cite{ego,ENOV4}, as discussed in more detail later. We also consider an extended range of reheating temperatures ranging from the Big-Bang Nucleosynthesis (BBN) limit, 4 MeV \cite{tr4} to instantaneous reheating, $\sim 2 \times 10^{15}$~GeV. We extend our analysis to generalized attractor models with monomial potentials $V(\varphi) \propto \varphi^k$ near the minimum, demonstrating that models with $k \geq 6$ can accommodate more easily the ACT preference for larger $n_s$. It has also been shown that slightly deformed potentials~\cite{accid}, that occur quite naturally when the inflation model is embedded in UV completion such as an SU(5) or SO(10) GUT theory \cite{deform}, may have a significant impact on the CMB observables, as we discuss in more detail below.

The remainder of this paper is organized as follows. In Section~\ref{sec:no-scale} we introduce and review no-scale models of inflation. In Section~\ref{sec:obs}, we compile and analyze the current observational constraints on $(n_s, r)$ from various CMB experiments. In Section~\ref{sec:infdynamics}, we present the inflationary dynamics and derive CMB observables. In Section~\ref{sec:infact}, we analyze generalized $\alpha$-attractors with higher-order minima as well as deformed models. In Section~\ref{sec:reheating}, we discuss the reheating constraints we use to limit $N_*$. We present the results in Section~\ref{sec:results} and conclude in Section~\ref{sec:conclusions}.

\section{Review of No-Scale Supergravity Models of Inflation}
\label{sec:no-scale}

There is a deep correspondence between the $R+R^2$ theory and no-scale supergravity \cite{ENO6,ENO7,DLT,DGKLP,eno9,building,Ema:2024sit}, and several well-motivated scenarios arising naturally in supergravity and string theory appear within the $\alpha$-attractor framework. The simplest no-scale supergravity \cite{no-scale,LN} models are characterized by a K\"ahler potential of the form 
\beq
K \; = \; -3 \, \alpha \, M_P^2 \, \ln\left(\frac{T + \bar{T}}{M_P} - \frac{|\phi|^2}{3M_P^2}\right) \, ,
\label{n-sK}
\eeq
where $\alpha = 1$ in the simplest model~\cite{ENO6}, but $\alpha \ne 1$ is possible in generalized models~\cite{ENO7}. Here, $T$ is the volume modulus and $\phi$ is a matter-like field. Exploiting the underlying SU(2,1)/SU(2)$\times$U(1) symmetry, the Starobinsky potential can be derived from many different superpotentials \cite{ENO6,ENO7,enov1}. For example, when $\alpha = 1$
\beq
W \; = \; M \left(\frac12 \phi^2 - \frac{1}{3\sqrt{3} M_P}\phi^3 \right) \, ,
\label{wz}
\eeq
leads to the Starobinsky potential \cite{ENO6} with the canonically-normalized inflaton related to $\phi$ through
\beq
\label{phi}
\phi \; = \; \sqrt{3} M_P \tanh \left(\frac{\varphi}{\sqrt{6}M_P} \right) \, .
\eeq
Alternatively (also when $\alpha = 1$), 
\begin{equation}
W \; = \sqrt{3} M M_P \phi \left(\frac{T}{M_P} - \frac{1}{2}\right)
\label{W3}
\end{equation}
leads to the Starobinsky potential \cite{Cecotti,ENO7} when the canonically-normalized inflaton is related to $T$ through
\beq
T = \frac{M_P}{2} e^{\sqrt{\frac23}\frac{\varphi}{M_P}}
\label{canT}
\, .
\eeq
The $\alpha$-attractor variations of the Starobinsky model can be derived in no-scale supergravity as discussed in more detail in Appendix \ref{appA}. Such variations arise naturally in compactified string models \cite{Witten}, where the volume modulus $T$ factorizes as a product of three independent moduli $T_i$. Models where inflation is driven by one (two) of these moduli correspond to $\alpha = 1/3$ (2/3)~\cite{ENO7}, while larger values of $\alpha$ can arise from complex structure moduli~\cite{Kallosh:2013hoa}.
The parameter $\alpha$ that determines the (constant) curvature of the K\"ahler manifold plays a crucial role in determining the inflationary predictions: larger values of $\alpha$ reduce the flatness of the potential plateau at the horizon-crossing field value $\varphi_*$, thereby modifying the observables $n_s$ and $r$. 
Generalizing the coefficient of the log in the K\"ahler potential to $\alpha \ne 1$ modifies the prediction for $r$ by a factor $\alpha$, as first pointed out in~\cite{ENO7} and subsequently in~\cite{KLR}.
Remarkably, as demonstrated in~\cite{ENO7,KLR,T-model,rs,ENOV3,building}, the predictions of all ($\alpha$-)attractor models converge to universal values in the large-$N_*$ limit:
\begin{equation}
\label{eq:cmbpredictions}
n_s \; \simeq \; 1 - \frac{2}{N_*}\,, \qquad r \; \simeq\; \frac{12\alpha}{N_*^2}\,,
\end{equation}
where the relation between $\varphi_*$ and $N_*$ is discussed in greater detail in Section \ref{sec:infdynamics} below.

This universality arises because the $\alpha$-Starobinsky (E-model) and T-model potentials have identical leading-order behaviors in their small-field expansions.~\footnote{The potentials~(\ref{eq:emodel}) and~(\ref{eq:tmodel}) agree at first order in $\exp(-\sqrt{2/(3\alpha)}\varphi/M_P)$ but differ at higher orders, yielding model-dependent corrections when $\varphi \sim \sqrt{\alpha}M_P$. These corrections become negligible for $N_* \gg 1$.} For the canonical Starobinsky model with $\alpha = 1$, the predicted values $n_s \simeq 0.961 - 0.968$ and $r \simeq 0.004 - 0.003$ (for $N_* = 50 - 60$ $e$-folds) were long considered a triumph of the theory, aligning perfectly with pre-ACT CMB measurements. However, these predictions now fall significantly below the P-ACT-LB2 central value of $n_s = 0.9752$, potentially signaling a departure from the previously established concordance.

A comprehensive analysis of $\alpha$-attractor models  using BICEP/Keck 2018 data was performed in~\cite{Ellis:2021kad}, which found that both $\alpha$-Starobinsky and T-models remained comfortably within the observational bounds available at that time. The study demonstrated that $\alpha$-Starobinsky models with $\alpha \in (0.67, 12)$ and T-models with $\alpha \in (1.3, 5.1)$ fell within the 68\% confidence region, while considering gravitino and dark matter constraints from supergravity. Prior to ACT DR6, these values sat comfortably within the confidence regions of all major CMB experiments, establishing Starobinsky inflation as a benchmark model.
For a related study see Ref.~\cite{Iacconi:2023mnw}.

\section{Observational Constraints}
\label{sec:obs}
Multiple CMB experiments have provided increasingly precise measurements of the scalar spectral index $n_s$ and stringent upper limits on the tensor-to-scalar ratio $r$. Table~\ref{tab:nsr} summarizes the current constraints from key datasets.

\begin{table}[h!]
\centering
\setlength{\tabcolsep}{4pt}      
\renewcommand{\arraystretch}{1.15}
\caption{Recent CMB constraints on the scalar spectral index $n_s$ (68\% CL) and the tensor-to-scalar ratio $r$ (95\% CL). Unless otherwise noted, values are quoted at a pivot scale $k_{*}=0.05\,\mathrm{Mpc}^{-1}$.}
\label{tab:nsr}
\begin{tabular*}{\columnwidth}{@{\hspace{0pt}}p{0.49\columnwidth}@{\extracolsep{\fill}}c c}
\hline\hline
Dataset & $n_s$ & $r$ \\ \hline
{\it Planck} 2018           & $0.9649 \pm 0.0044$ & $<0.11^{\,b}$ \\
(TT,TE,EE+lowE)$^a$   &                     &               \\
{\it Planck} 2018 + lensing$^a$                & $0.9649 \pm 0.0042$ & $<0.10^{\,b}$ \\
{\it Planck} 2018 + lensing                    & $0.9665 \pm 0.0038$ & $<0.11^{\,b}$ \\
+ BAO$^a$                                &                     &                \\
{\it Planck} 2018 + BK15$^a$                   & $0.9668 \pm 0.0037$ & $<0.058^{\,b}$ \\
+ lensing + BAO$^a$                                &                     &                \\          
{\it Planck} 2018 + BK18$^{c}$                   & — & $<0.036$ \\
\hline
ACT DR6$^d$                    
& $0.9666 \pm 0.0077$ & — \\
ACT DR6 + WMAP$^d$                       & $0.9660 \pm 0.0046$ & — \\
{\it Planck} + ACT DR6$^d$                     & $0.9709 \pm 0.0038$ & — \\
{\it Planck} + ACT DR6                         & $0.9752 \pm 0.0030$ & — \\
+ lensing + BAO$^d$                      &                     &   \\
{\it Planck} + ACT DR6 + BK18$^e$              & —                   & $<0.038$ \\
\hline
SPT-3G (2019-2020)$^f$                  & —                   & $<0.25$ \\
SPT-3G + {\it Planck}$^g$                     & $0.9647 \pm 0.0037$ & — \\
SPT-3G D1$^h$                           & $0.951 \pm 0.011$ & — \\
SPT-3G D1+ ACT$^h$                      & $0.9671 \pm 0.0058$ & — \\
SPT-3G D1 + {\it Planck}$^h$                  & $0.9636 \pm 0.0035$ & — \\
CMB-SPA$^h$                             & $0.9684 \pm 0.0030$ & — \\
\hline\hline
\end{tabular*}
\flushleft\footnotesize
$^{a}$\cite{Planck}; 
$^{b}$At a pivot scale $k_{*}=0.002\,\mathrm{Mpc}^{-1}$; 
$^{c}$\cite{BICEP2021}; 
$^{d}$\cite{ACT:2025fju}; 
$^{e}$\cite{ACT:2025tim}; 
$^{f}$\cite{SPT-3G:2025vtb};
$^{g}$\cite{SPT-3G:2024atg};
$^{h}${\cite{SPT-3G:2025bzu}}
\end{table}

\subsection*{\textbf{\textit{Planck} 2018}}
The \textit{Planck} satellite 2018 data release~\cite{Planck} remains the baseline for CMB parameter constraints. The analysis of temperature and polarization data yields $n_s = 0.9649 \pm 0.0044$ (68\% CL) from the TT, TE, EE + low E + likelihood, with negligible change when including CMB lensing as given in Eq.~(\ref{nsexp}). This is an $8\sigma$ detection of a red-tilted spectrum ($n_s < 1$) and definitively rules out exact scale invariance. For tensor modes, \textit{Planck} alone constrains $r_{0.002} < 0.11$ (95\% CL).
This limit was strengthened by the BICEP/Keck 2015 result of $r < 0.062$. 
When combined with BICEP/Keck 2015 data, the {\it Planck} result tightens to $r_{0.002} < 0.058$ (95\% CL) with a slight shift in the spectral index to $n_s = 0.9668 \pm 0.0037$~\cite{Planck}. These constraints exclude simple polynomial potentials such as $V(\phi) \propto \phi^4$ at more than $3\sigma$ and favor models with plateau-like potentials, particularly those with $V''(\phi) < 0$ during inflation.

\subsection*{BICEP/Keck 2018}
The BICEP/Keck collaboration targets degree-scale $B$-mode polarization from a dedicated South Pole observatory, providing the most stringent constraints on inflationary gravitational waves. The progression of limits illustrates the rapid experimental progress: BICEP2+Keck+\textit{Planck} (data up to 2014) constrained $r_{0.05} < 0.09$ (95\% CL) \cite{Planck:2015fie}, while BICEP/Keck data up to 2015 (BK15) combined with \textit{Planck} constrained $r_{0.05} < 0.062$ (95\% CL)~\cite{BICEP2:2018kqh}. The latest analysis incorporating BICEP3 and Keck Array observations up to 2018 (BK18) yields the current best limit: $r_{0.05} < 0.036$ (95\% CL)~\cite{BICEP2021},~\footnote{A subsequent analysis in~\cite{Tristram:2021tvh} used BB autocorrelation data from \cite{PR4} and obtained $r_{0.05} < 0.032$, allowing for a free reionization optical depth.} and a likelihood analysis yielded $r_{0.05} = 0.014^{+0.010}_{-0.011}$. These constraints assume the inflationary consistency relation $n_t = -r/8$. 
The constraint corresponds to an uncertainty $\sigma(r_{0.05}) = 0.009$, approaching the sensitivity needed to detect $r \sim 0.01$ models. While BICEP/Keck observations do not significantly constrain $n_s$ (which is determined by large-scale temperature and $E$-mode data), they provide crucial complementarity to \textit{Planck}.

\subsection*{ACT DR6}
The ACT Data Release 6 produced high-resolution maps of the CMB temperature and polarization measurements over 40\% of the sky, complementing \textit{Planck}'s full-sky coverage with superior small-scale sensitivity~\cite{ACT:2025fju,ACT:2025tim}. ACT's arcmin resolution probes small-scale multipoles complementary to \textit{Planck}. These data have been combined with {\it Planck} and WMAP in order to extend down to low multipoles and are in good agreement with both {\it Planck} and WMAP  larger-scale data. For example, the combination of WMAP and ACT (W-ACT) yielded~\footnote{Unless specifically noted all uncertainties are given as 68\% CL. } $n_s = 0.9660 \pm 0.0046$, whereas {\it Planck} without lensing data give $n_s = 0.9649 \pm 0.0044$. However, because the large-scale data result in a positive correlation between $n_s$ and the baryon density $\Omega_{\rm b} h^2$, in contrast to them being anti-correlated in the ACT data, and since the W-ACT determination of the baryon density is
slightly higher than that of {\it Planck} alone ($\Omega_{\rm b} h^2 = 0.02263 \pm 0.00012$ vs $0.02237 \pm 0.00015$), the combination of {\it Planck} and ACT (P-ACT) tends to increase the fit value of $n_s = 0.9709 \pm 0.0038$ and P-ACT with lensing gives $0.9713 \pm 0.0037$, about $1.5$--$2\sigma$ higher than the {\it Planck} result. 

ACT has also combined their data with DESI DR1 data~\cite{DESI:2024uvr, DESI:2024mwx} and subsequently DR2 data~\cite{DESI:2025zpo,DESI:2025zgx}. These combinations come with potential additional uncertainties due to unknown systematics in the supernovae data (see, e.g., \cite{dovekie}). This combination has yielded new CMB constraints that support inflation broadly while further challenging specific models. Most notably, these data imply a higher baryon density of $\Omega_{\rm b} h^2 = 0.02256 \pm 0.00011$ for the {\it Planck}-ACT-DESI DR1 (P-ACT-LB) combination of data and $\Omega_{\rm b} h^2 = 0.02258 \pm 0.00010$ using DESI DR2 (P-ACT-LB2). This shifts the scalar spectral index $n_s$ significantly upward from previous measurements. 
The combination of \textit{Planck} + lensing, ACT, and DESI BAO (P-ACT-LB) pushes the tilt to $n_s = 0.9743 \pm 0.0034$ and  even higher  using DR2 data (P-ACT-LB2)
\cite{ACT:2025fju}
\beq
n_s = 0.9752 \pm 0.0030~~~ (68\% {\rm~CL}) \, .
\label{nsact}
\eeq
 The P-ACT-LB2 results lie approximately $2\sigma$ above the {\it Planck} 2018 constraint, representing a significant shift in their preferred value of $n_s$.
Some caution is warranted, however, as the shift may be reflective of underlying tension between BAO/CMB analyses \cite{Ferreira:2025lrd}.

For tensor constraints, the ACT DR6 measurements, which favor $n_s \simeq 0.975$, lack direct sensitivity to degree-scale $B$-modes and rely on the BICEP/Keck likelihood. The combined \textit{Planck}+ACT+BK18 analysis maintains $r < 0.038$ (95\% CL)~\cite{ACT:2025fju}, with the modest relaxation from BK18 alone reflecting parameter degeneracies in the expanded dataset. This combination of $n_s$ and $r$ places the canonical Starobinsky model ($\alpha = 1$) outside the 95\% confidence region, indicating a $\sim 2\sigma$ tension with its theoretical predictions~\cite{ACT:2025tim}. 
At $ r=0.004$, the P-ACT-LB 95 \% CL range on $n_s$ is $0.967 < n_s < 0.982$. 
As noted above, this result merits careful consideration, particularly given the variation among more recent CMB analyses.

\subsection*{SPT-3G}
The third-generation South Pole Telescope camera (SPT-3G) represents a major upgrade in polarization sensitivity and sky coverage. 
Initial results from the 2019-2020 observing seasons demonstrate the instrument's capabilities. SPT-3G measures the $B$-mode power spectrum over $30 < \ell < 500$, achieving $r < 0.25$ (95\% CL) \cite{SPT-3G:2025vtb} from these data alone. While this constraint is weaker than BICEP/Keck results, it represents only 5\% of the planned dataset and serves primarily as a systematic check and foreground characterization.

While SPT alone does not determine $n_s$ at high precision (the uncertainty is $0.011$), 
the combination with \textit{Planck} yields competitive cosmological constraints: $n_s = 0.9647 \pm 0.0037$ (68\% CL) from SPT+\textit{Planck}~\cite{SPT-3G:2024atg}, consistent with the original \textit{Planck} value and in tension with ACT DR6.
The latest SPT-3G analysis incorporating data through 2020 confirms a similar result, with $n_s \simeq 0.9636 \pm 0.0035$ in combination with \textit{Planck}~\cite{SPT-3G:2025bzu}. 
This is also significantly lower than the P-ACT value.

 The combined data sets of SPT+\textit{Planck}+ACT (denoted CMB-SPA) yield~\cite{SPT-3G:2025bzu} the following spectral tilt:
\beq
n_s = 0.9684 \pm 0.0030 ~~~ (68\%~{\rm CL}) \, ,
\label{nsspa}
\eeq 
which is consistent with the \textit{Planck} result (at 1$\sigma$).
Adding DESI BAO results in $n_s = 0.9728 \pm 0.0027$, somewhat lower than the P-ACT-LB result. 
While further data (e.g., extended DESI analyses and SPT-3G observations) will help clarify these differences, the current tension between the ACT-shifted constraints and standard Starobinsky model predictions requires theoretical examination.

However, a new 2.8$\sigma$ discrepancy has emerged in the $\Lambda$CDM between the combined CMB experiments and DESI BAO measurements, suggesting potential tensions in our understanding of late-time cosmology. This discrepancy between contemporary ground-based experiments and BAO data highlights the importance of systematic cross-checks and motivates exploration of extended cosmological models. Looking ahead, SPT-3G aims for a tensor sensitivity $\sigma(r) \sim 0.003$ through delensing techniques and five years of integrated observations, potentially accessing $r \sim 0.01$ inflationary models. The expanded SPT-3G survey will eventually cover 25\% of the sky, dramatically improving constraints and enabling more stringent tests of cosmological models.

In what follows, we compare the predictions of models of inflation with the three CMB results that are highlighted in Eqs.~(\ref{nsexp}), (\ref{nsact}), and (\ref{nsspa}). Note that only the P-ACT-LB results make use of BAO data. 

\section{Inflationary Dynamics, \\
CMB Observables and Reheating}
\label{sec:infdynamics}
The dynamics of the inflaton field $\varphi$ is governed by the action
\begin{equation}    
    \label{eq:action}
    \mathcal{S} \; = \; \int d^4 x \sqrt{-g} \left[-\frac{M_P^2}{2} R + \frac{1}{2} g^{\mu\nu}\partial_{\mu} \varphi \partial_{\nu} \varphi - V(\varphi) \right] \,,
\end{equation}
where $V(\varphi)$ is the inflaton potential given by Eq.~(\ref{eq:emodel}) for the $\alpha$-Starobinsky (E-model) or Eq.~(\ref{eq:tmodel}) for the T-model. In the slow-roll approximation, the inflationary dynamics is characterized by the slow-roll parameters:
\begin{equation}
\label{eq:epseta}
\varepsilon_V \; \equiv \; \frac{M_{P}^{2}}{2}\left(\frac{V'}{V}\right)^{2}\,, \qquad \eta_V \; \equiv \; M_{P}^{2}\frac{V''}{V} \,,
\end{equation}
where primes denote derivatives with respect to $\varphi$. During slow-roll inflation, these parameters satisfy $\varepsilon_V, |\eta_V| \ll 1$. The number of $e$-folds from horizon exit of the pivot scale to the end of inflation is given by
\begin{equation}
\begin{aligned}
\label{eq:efolds}
N_{*} \; = \; \int_{t_*}^{t_{\rm end}} H \, dt \; \simeq \;& \frac{1}{M_{P}^{2}} \int_{\varphi_{*}}^{\varphi_{\mathrm{end}}} \frac{V(\varphi)}{V'(\varphi)} d \varphi  \\
\; \simeq \;& \int_{\varphi_{\mathrm{end}}}^{\varphi_{*}} \frac{1}{\sqrt{2 \varepsilon_V}} \frac{d \varphi}{M_{P}}\,,
\end{aligned}
\end{equation}
where $\varphi_*$ denotes the field value when the pivot scale $k_*$ exits the horizon. The \textit{Planck} collaboration adopted $k_* = 0.05 \, \mathrm{Mpc}^{-1}$, and we also adopt this except if stated otherwise. The end of inflation is defined by the condition $\ddot{a} = 0$, where $a(t)$ is the scale factor corresponding to $\varepsilon_H(\varphi_{\rm end}) = 1$, and $\varepsilon_H \equiv (M_P^2/2)(H'/H)^2$. In terms of $\varepsilon_V$, inflation ends when \cite{EGNO5}
\beq
 \varepsilon_V \; \simeq \; (1 + \sqrt{1 - \eta_V/2})^2 \, .
 \label{epsV}
\eeq
The primary CMB observables are expressed in terms of the slow-roll parameters evaluated at horizon exit:
\begin{align}
    \label{eq:spectrtilt}
    n_{s} &\simeq 1 - 6 \varepsilon_{V*} + 2 \eta_{V*} \, , \\
    \label{eq:sclrtotens}
    r & \simeq 16 \varepsilon_{V*} \,, \\
    \label{eq:powerspectr}
    A_{s} &= \frac{V_{*}}{24 \pi^{2} \varepsilon_{V*} M_{P}^{4}} \,,
\end{align}
where the asterisk indicates quantities evaluated at $\varphi_*$. The amplitude of scalar perturbations is observationally constrained to $\ln(10^{10} A_s) = 3.044 \pm 0.014$ at 68\% CL from \textit{Planck} 2018 TT,TE,EE+lowE+lensing data~\cite{Planck}, corresponding to $A_s \simeq 2.10 \times 10^{-9}$. For the Starobinsky potential Eq.~(\ref{eq:powerspectr}) is equivalent to Eq.~(\ref{AsStaro}).

For the attractor models described by Eqs.~(\ref{eq:emodel}) and~(\ref{eq:tmodel}), the predictions in the large-$N_*$ limit take the particularly simple form given by Eq.~(\ref{eq:cmbpredictions})~\cite{ENO7}. These expressions are valid for $\alpha \lesssim \mathcal{O}(1)$ and $N_* \gg 1$. For the Starobinsky limit ($\alpha = 1$), more precise analytical expressions including higher-order corrections can be found in Ref.~\cite{ENOV4}. We note that the prediction for $n_s$ is independent of $\alpha$ to leading order, making it a robust prediction of this class of models.

Using the $e$-fold expression~(\ref{eq:efolds}), we can derive the inflaton field value at horizon crossing for the pivot scale $k_*$. For the case $\alpha = 1$, the field values are given by~\cite{EGNO5}:
\begin{align}
\label{alphastarophistar}
\frac{\varphi_{*}}{M_P} &= \sqrt{\frac{3}{2}} \left[1 + \frac{3}{4N_* - 3}\right] \nonumber \\
&\times \ln\left(\frac{4N_*}{3} + e^{\sqrt{\frac{2}{3}}\frac{\varphi_{\rm end}}{M_P}} - \sqrt{\frac{2}{3}}\frac{\varphi_{\rm end}}{M_P}\right) \,, \\
\label{alphastarophiend}
\frac{\varphi_{\rm end}}{M_P} &= \sqrt{\frac{3}{2}} \ln\left[\frac{2}{11}(4 + 3\sqrt{3})\right] \; \simeq \; 0.63 \,, \\
& \qquad \qquad \qquad \qquad \qquad(\text{$\alpha$-Starobinsky (E-Model))}\nonumber \, ,
\end{align}
\begin{align}
\label{tmodelphistar}
\frac{\varphi_{*}}{M_P} &= \sqrt{\frac{3}{2}} \cosh^{-1}\left(\frac{4N_*}{3} + \cosh\left(\sqrt{\frac{2}{3}}\frac{\varphi_{\rm end}}{M_P}\right)\right), \\
\label{tmodelphiend}
\frac{\varphi_{\rm end}}{M_P} &= \sqrt{\frac{3}{2}} \ln\left[\frac{1}{11}(14 + 5\sqrt{3})\right] \; \simeq \; 0.89 \,, \\
& \qquad \qquad \qquad \qquad \qquad \qquad \qquad (\text{T-Model}) \nonumber \, . 
\end{align}
Full analytical expressions including the dependence on $\alpha$ are provided in Appendix~\ref{appB}, Eqs.~(\ref{eq:phiendE})-(\ref{eq:phistarT}).

We next calculate the number of $e$-folds $N_*$ between horizon exit and the end of inflation, accounting for the post-inflationary evolution. Assuming no additional entropy production between the end of reheating and horizon re-entry, we have~\cite{Martin:2010kz, LiddleLeach}:
\begin{equation}
\label{eq:nstarreh}
\begin{aligned}
N_{*} &= \ln\left[\frac{1}{\sqrt{3}}\left(\frac{\pi^{2}}{30}\right)^{1/4}\left(\frac{43}{11}\right)^{1/3} \frac{T_{0}}{H_{0}}\right] - \ln\left(\frac{k_{*}}{a_{0}H_{0}}\right) \\
&\quad - \frac{1}{12}\ln g_{\mathrm{RH}} + \frac{1}{4}\ln\left(\frac{V_{*}^{2}}{M_{P}^{4}\rho_{\mathrm{end}}}\right) \\
&\quad + \frac{1-3w_{\mathrm{int}}}{12(1+w_{\mathrm{int}})}\ln\left(\frac{\rho_{\mathrm{rad}}}{\rho_{\mathrm{end}}}\right) \,,
\end{aligned}
\end{equation}
where $H_0 = 67.36$ km s$^{-1}$ Mpc$^{-1}$~\cite{Planck} and $T_0 = 2.7255$ K~\cite{Fixsen:2009ug} are the present-day Hubble parameter determined by {\it Planck} and the CMB temperature, respectively. Here, $\rho_{\mathrm{end}} = 3M_P^2H_{\mathrm{end}}^2$ is the energy density at the end of inflation, and $\rho_{\rm rad}$ is the energy density in radiation in the radiation-dominated era  when $w = p/\rho = 1/3$. Up to a sub-percent correction in $N_*$, this energy density can be associated with the energy density at the end of reheating, $\rho_{\rm RH}$, defined when energy densities in radiation and the inflaton are equal,  $\rho_\varphi = \rho_{\rm R}\equiv \rho_{\rm RH}$. In turn, this can be related to the reheating temperature $T_{\rm RH}$ via the thermodynamic relation $ \rho_{\rm RH} =  (g_{\rm RH} \pi^2/30) T_{\rm RH}^4$, where $g_{\mathrm{RH}}$ denotes the number of effective relativistic degrees of freedom at reheating, which is $g_{\rm{reh}} =  915/4$ in the minimal supersymmetric extension of the Standard Model (MSSM), and $a_0 = 1$ is the scale factor at the present day. The equation of state parameter during reheating is characterized by its average:
\begin{equation}
w_{\mathrm{int}} \equiv \frac{1}{N_{\mathrm{rad}}-N_{\mathrm{end}}} \int_{N_{\mathrm{end}}}^{N_{\mathrm{rad}}} w(n) \, d n \, .
\end{equation}
For a detailed discussion on the computation of $N_*$, see Appendix~\ref{appB}.
For the \textit{Planck} pivot scale $k_* = 0.05$ Mpc$^{-1}$, the first two terms in Eq.~(\ref{eq:nstarreh}) yield:
\begin{equation}
N_* \; = \; 61.49 - \frac{1}{12}\ln g_{\mathrm{RH}} + \Delta N_{\mathrm{inf}} + \Delta N_{\mathrm{rad}} \,,
\end{equation}
where $\Delta N_{\mathrm{inf}}$ and $\Delta N_{\mathrm{rad}}$ represent the model-dependent contributions from inflation and reheating, respectively.

We consider three benchmark reheating scenarios:

(i) MSSM reheating: Taking $g_{\mathrm{RH}} = 915/4$ corresponding to the MSSM degrees of freedom, we obtain
\begin{equation}
N_*^{\mathrm{MSSM}} = 61.04 + \cdots \, .
\end{equation}

(ii) Electroweak-scale reheating: For $T_{\mathrm{reh}} = 100$ GeV with SM degrees of freedom ($g_{\mathrm{RH}} = 427/4$), we find
\begin{equation}
N_*^{\mathrm{EW}} = 61.10 + \cdots \, .
\end{equation}

(iii) BBN compatible reheating: The minimum temperature consistent with BBN is $T_{\mathrm{reh}} \gtrsim \mathcal{O}(1)$ MeV. Taking $T_{\mathrm{BBN}} = 4$ MeV with $g_{\mathrm{RH}} = 10.75$, we obtain
\begin{equation}
N_*^{\mathrm{BBN}} = 61.29 + \cdots \, .
\end{equation}
As we shall see, typical values of $N_* \simeq 45 - 55$ encompasses these scenarios for typical inflationary potentials and reheating dynamics with $-1/3 \leq w_{\mathrm{int}} \leq 1$.

To determine $N_*$ numerically for different reheating scenarios, we solve the coupled system of equations governing the post-inflationary dynamics. The evolution of the inflaton and radiation energy densities during reheating is described by:
\begin{align}
\label{eq:dyn1}
\dot{\rho}_{\varphi} + 3H(1 + w_\varphi)\rho_{\varphi} &= -\Gamma_{\varphi}\rho_{\varphi} \,, \\
\label{eq:dyn2}
\dot{\rho}_{r} + 4H\rho_{r} &= \Gamma_{\varphi}\rho_{\varphi} \,, \\
\label{eq:dyn3}
\rho_{\varphi}+\rho_{r} &=3 M_{P}^{2} H^{2} \, ,
\end{align}
where $\Gamma_{\varphi}$ is the inflaton decay rate, and $w_\varphi$ is the inflaton equation of state parameter. The time-averaged equation of state during reheating evolves according to:
\begin{equation}
\label{eq:dyn4}
\frac{d}{dt}(Nw_{\mathrm{int}}) = Hw \, ,
\end{equation}
where $w = (\rho_\varphi w_\varphi + \rho_r/3)/(\rho_\varphi + \rho_r)$ is the instantaneous total equation of state.

We assume that reheating occurs through the perturbative decay of the inflaton to either 
scalars or fermions. For potentials with an approximate quadratic minimum (as in Eqs.~(\ref{eq:emodel}) and (\ref{eq:tmodel})) the reheating temperature is determined by the decay rate, $\Gamma_\varphi$,
\beq
\frac{g_{\rm RH} \pi^2}{30} T_{\rm RH}^4  \; = \; \frac{12}{25} \Gamma_\varphi^2 M_P^2 \, .
\label{trh}
\eeq
For the generalized models discussed in the next Section, the corresponding expressions for $T_{\rm RH}$ in terms of the decay rate have been derived in Ref. \cite{gkmo2}. We allow the reheating temperature to take values as low as 4 MeV to ensure standard BBN \cite{tr4} and in principle
as high as the perturbative limit at $T_{\rm RH} \sim 10^{15}$~GeV. A more restrictive range is $100 \lesssim T_{\rm RH}/{\rm GeV} \lesssim 10^{10}$, where the lower limit would allow for leptogenesis such that sphaleron interactions reach equilibrium \cite{fy} and the upper limit could be imposed in supersymmetric models to avoid the overproduction of a 100 GeV gravitino and supersymmetric dark matter~\cite{ego,ENOV4}.

Given the value of $\varphi_*$, determined from Eqs.~(\ref{eq:efolds}) and (\ref{eq:nstarreh}), we can 
 combine these field values with the amplitude constraint from Eq.~(\ref{eq:powerspectr}) to obtain the normalization of the inflaton potential, which to a good approximation is given by
\begin{equation}
\label{eq:infnorm}
\lambda \; \simeq \; \frac{24\alpha\pi^2 A_{s}}{N_*^2} \, ,
\end{equation}
where $A_s \simeq 2.1 \times 10^{-9}$. This demonstrates that the energy scale of inflation is proportional to $\sqrt{\alpha}$, with larger $\alpha$ corresponding to higher inflationary energy scales. For the canonical Starobinsky model ($\alpha = 1$, $N_* = 55$), this yields $\lambda \simeq 1.6 \times 10^{-10}$, corresponding to an inflationary Hubble scale $H_* \simeq 1.5 \times 10^{13}$ GeV. 

\section{Inflationary Attractors}
\label{sec:infact}

\subsection{Generalized Attractor Models}
\label{sec:generalized}

 The predictions of the canonical E- and T-models (\ref{eq:emodel},\ref{eq:tmodel}) which yield $n_s \simeq 0.957 - 0.963$ for reheating temperatures between $100-10^{10}$~GeV, as we demonstrate in Section~\ref{sec:results}, are- significantly below the value of $n_s \simeq 0.975$ preferred by the ACT DR6 data, and motivates consideration of generalized attractor potentials~\cite{Carrasco:2015rva, ENOV3, ENOV4}:
\begin{align}
\label{eq:alphastarogeneral}
V & \; = \; \frac{3}{4} \lambda  M_{P}^{4}\left(1-e^{-\sqrt{\frac{2}{3 \alpha}} \frac{\varphi}{M_{P}}}\right)^{k} \, , & \nonumber \\
& \; \; (\text{Generalized $\alpha$-Starobinsky (E-model)})&  \\
\label{eq:tmodelgeneral}
V & \; = \; \frac{3}{4}  \lambda M_P^4 \tanh^{k} \left(\frac{\varphi}{\sqrt{6 \alpha}M_P } \right) \,, & \nonumber \\
& \; \; (\text{Generalized T-Model}) &
\end{align}
where $k \geq 2$ is an even integer. These potentials retain the attractor behavior at large field values while modifying the approach to the minimum, with $V(\varphi) \propto \varphi^k$ for small $\varphi$. As a consequence, the inflaton mass near the minimum is given by:
\begin{align}
\label{eq:infmass1gen}
m_{\varphi}^{\text{(E)}} &= \frac{\sqrt{3}}{2} \sqrt{\left(\frac{2}{3}\right)^{\frac{k}{2}}k(k-1) \alpha^{-\frac{k}{2}} \lambda \varphi^{k-2}} \, ,
\\
\label{eq:infmass2gen}
m_{\varphi}^{\text{(T)}} &= \frac{\sqrt{3}}{2} \sqrt{6^{-\frac{k}{2}}k(k-1) \alpha^{-\frac{k}{2}} \lambda \varphi^{k-2}} \,.
\end{align}
For all $k>2$, the inflaton mass \emph{vanishes} at the minimum as $\varphi^{(k-2)/2} \to 0$, but these expressions reduce to the more familiar non-zero masses for the quadratic case ($k=2$): $m_\varphi^{(E)}=\sqrt{\lambda/\alpha}$ and $m_\varphi^{(T)}=\frac{1}{2}\sqrt{\lambda/\alpha}$. 

One can show that in the large-$N_*$ limit, the generalized E-models predict: 
\begin{align}
\label{eq:genemodns}
n_s& \; \simeq \; 1 - \frac{2}{N_*} - \frac{3 \alpha(2 + k - 2\log[\frac{2kN_*}{3 \alpha}])}{2k N_*^2} \, ,   \\
\label{eq:genemodr}
r & \; \simeq \; \frac{12 \alpha}{N_*^2} - \frac{36 \alpha^2 \log[\frac{2kN_*}{3 \alpha}]}{k N_*^3} \, ,
\end{align}
and the generalized T-models predict
\begin{align}
\label{eq:gentmodns}
n_s& \; \simeq \; 1 - \frac{2}{N_*} - \frac{3(k - 2) \alpha}{2kN_*^2} \, ,   \\
\label{eq:gentmodr}
r & \; \simeq \; \frac{12 \alpha}{N_*^2} - \frac{36 \alpha^2}{k N_*^3} \, .
\end{align}
Note that the normalization of $\lambda$ in Eq.~(\ref{eq:infnorm}) remains independent of $k$. Generalizations of the analytical expressions for $\varphi_{\rm end}$ and $\varphi_*$ including the dependence on both $\alpha$ and $k$ are provided in Appendix~\ref{appB}, Eqs.~(\ref{eq:phiendEk})-(\ref{eq:phistarTk}).

\begin{table}[ht!]
\centering
\setlength{\tabcolsep}{10pt}
\renewcommand{\arraystretch}{1.2}
\begin{tabular}{c|cc|cc}
\hline
\multirow{2}{*}{$k$} & \multicolumn{2}{c|}{E-model} & \multicolumn{2}{c}{T-model} \\
                     & $n_s$        & $r$           & $n_s$        & $r$          \\
\hline
2 & 0.9649 & 0.00355 & 0.9638 & 0.00387\\
4  & 0.9643 & 0.00371 & 0.9636 & 0.00390 \\
6  & 0.9641 & 0.00378 & 0.9636 & 0.00391 \\
8  & 0.9639 & 0.00381 & 0.9636 & 0.00391 \\
10 & 0.9638 & 0.00383 & 0.9636 & 0.00391 \\
\hline
\end{tabular}
\caption{Predictions for the scalar spectral index $n_s$ and tensor-to-scalar ratio $r$ in E-models and T-models for $\alpha = 1$ and $N_* = 55$, as a function of the parameter $k$. }
\label{tab:EvsTmodels}
\end{table}

We summarize the values of the CMB observables $n_s$ and $r$ predicted for $N_* = 55$ $e$-folds and $\alpha = 1$ in Table~\ref{tab:EvsTmodels}. These values are exact, calculated by numerically solving for the primordial spectra, as discussed in Appendix~\ref{appC}. We observe that, for fixed $N_*$, $n_s$ decreases slightly as $k$ increases. However, the key insight is that models with $k > 4$ can accommodate a larger number of $e$-folds due to modified reheating dynamics. The equation of state parameter during reheating becomes $w_{\varphi} = (k-2)/(k+2)$, corresponding to a stiffer equation of state. This allows for higher values of $N_*$, which can more than compensate for the direct $k$-dependence and yield larger $n_s$ values.~\footnote{However, fragmentation effects during preheating impose an upper limit $N_{*, \, \rm{max}}$, which we discuss in Section~\ref{sec:results}.} Models with $k \geq 6$ can thus potentially provide better agreement with ACT DR6 constraints ($n_s \simeq 0.974$) while maintaining successful predictions for $r$, provided the allowed range of $N_*$ is sufficiently extended. We analyze these trade-offs in detail in Section~\ref{sec:results}.

These generalized attractors can arise naturally in no-scale supergravity.  The supergravity construction, detailed in Appendix~\ref{appA}, yields potentials of the form~(\ref{eq:alphastarogeneral}) and~(\ref{eq:tmodelgeneral}) through appropriate choices of the superpotential $W$.

\subsection{Deformed No-Scale Attractors}
\label{sec:noscaleattractors}
We can  further generalize the attractor models 
by adding in a deformation parameter $\kappa$. Such deformations lift the plateau nature of the potential at large field value and can have dramatic consequences for the CMB observables, most notably for $n_s$ \cite{ENO6,accid,deform}. These too can be
constructed within the no-scale supergravity framework. The model building aspects are detailed in Appendix~\ref{appA}.

We first consider the modified Starobinsky model with potential
\begin{equation}
\begin{aligned}
    \label{eq:modstaro}
    V&\; = \; \frac{3}{4} \lambda M_P^4 \left( \kappa -\kappa \cosh\left(\sqrt{\frac{2}{3\alpha}}\varphi \right) + \sinh \left(\sqrt{\frac{2}{3\alpha}} \varphi \right) \right)^k \, ,
\end{aligned}
\end{equation}
which reduces to the generalized $\alpha$-Starobinsky model~(\ref{eq:alphastarogeneral}) when $\kappa = 1$.~\footnote{The case $\alpha = 1, k = 2$ was considered in~\cite{ENO6} and more recently in~\cite{accid}, where $\kappa$ was denoted by $\lambda$.} For $\kappa$ slightly below unity, this modification introduces a small deviation from the standard attractor behavior while preserving the overall inflationary dynamics. In Section \ref{sec:results}, we examine a representative value: $\kappa = 0.9999$. This small deviation from unity generates observable effects that bring the predictions into excellent agreement with the ACT DR6 data, particularly by shifting the spectral index $n_s$ and tensor-to-scalar ratio $r$ along trajectories that intersect the observational contours~\cite{accid}.

\begin{figure*}[ht!]
    \centering
    \includegraphics[width=1\linewidth]{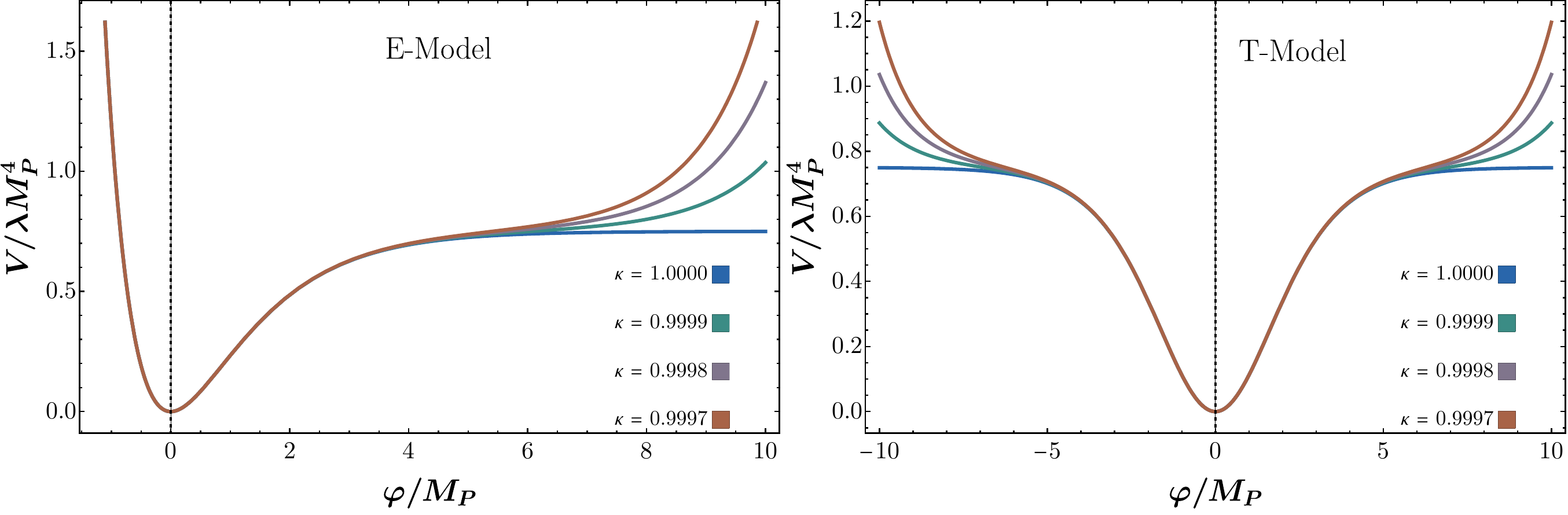}
    \caption{Deformed attractor potentials with $\alpha = 1$ and $k = 2$. Left panel: Deformed Starobinsky (E-model) potential for $\kappa = 1$, $0.9999$, $0.9998$, and $0.9997$. Right panel: Deformed T-model potential for the same $\kappa$ values. The deviations from $\kappa = 1$ introduce observable modifications to the inflationary plateau, and become substantial for $\varphi/M_P \gtrsim 6$.}
    \label{fig:attractors_kappa}
\end{figure*}

Similarly, one can construct modified T-model attractors with potential
\begin{equation}
\begin{aligned}
    \label{eq:modtmodel}
    V \; &= \; \frac{3}{16} \lambda M_P^4 \left( 
    1 + \kappa - (\kappa - 1) \cosh\left( \sqrt{\frac{2}{3\alpha}}\varphi \right)
    \right)^2\\
\; &\times \;\tanh^k\left( \frac{\varphi}{\sqrt{6\alpha}} \right) \, ,
\end{aligned}
\end{equation}
which reduces to the generalized T-model in Eq.~(\ref{eq:tmodel}) for $\kappa = 1$. 
As with the modified Starobinsky model, we investigate the same value of $\kappa = 0.9999$. These modifications preserve the attractor nature of the models while introducing controlled deviations that enhance compatibility with current observations. 

Fig.~\ref{fig:attractors_kappa} illustrates the evolution of the potential shapes for both model classes as $\kappa$ varies. The modifications introduce subtle but phenomenologically significant changes to the inflationary plateau region, which directly impact the predicted observables.

\section{Reheating Constraints and Dark Matter Production}
\label{sec:reheating}

Following inflation, for $k=2$ the Universe enters a matter-dominated phase as the inflaton begins to oscillate harmonically about its minimum. For larger $k$, the equation of state is $w = (k-2)/(k+2)$ and the oscillations are in general anharmonic.  The reheating process begins almost immediately after inflation ends. If the inflaton is allowed to decay into Standard Model particles with rate $\Gamma_\varphi$, the decay products thermalize \cite{Davidson:2000er,Harigaya:2013vwa,Mukaida:2015ria,GA,Passaglia:2021upk,Drees:2021lbm,Drees:2022vvn,Mukaida:2022bbo} and the temperature of this dilute plasma quickly reaches a maximum value~\cite{Giudice:2000ex, Ellis:2015jpg}:
\begin{equation}
T_{\max} =  \left[ \frac{30}{g_{\rm max} \pi^2} \frac{\sqrt{3}}{4} \left(\frac38 \right)^\frac35 \Gamma_{\varphi} \rho_{\rm end}^\frac12 M_P \right]^{1/4} \, ,
\end{equation}
for $k=2$.~\footnote{For more general expressions that
depend on $k$ and the spin of the final-state decay products, see Ref.~\cite{gkmo2}.} As the inflaton continues to decay, the temperature drops from its maximum as $T \propto a^{-3(k-1)/(2k+4)}$ for decays to fermions and $T \propto a^{-3/(2k+4)}$ for decays to scalars ($T \propto a^{-3/8}$ for $k=2$ for decays to both fermions and bosons) until reheating,~\footnote{If the reheating temperature is sufficiently low, the self-fragmentation of the inflaton will occur before the end of reheating, and the evolution of the instantaneous temperature depends on the decay of free inflaton quanta, see Section~\ref{sec:results}.} when the energy density in radiation starts to dominate the expansion, and subsequently $T \propto a^{-1}$ for all $k$ \cite{GKMO,gkmo2}. 
The reheating temperature for $k=2$ is given in Eq.~(\ref{trh}).

In the Starobinsky model derived from Eq.~(\ref{SStaro}), reheating is most naturally achieved by including the Standard Model (SM) in the action. Then, after the conformal transformation to the Einstein frame, there is a 
non-negligible coupling of the inflaton to the Higgs kinetic term leading to a decay rate,
$\Gamma_\varphi \propto M^3/M_P^2$, and a reheating temperature of $\simeq 3 \times 10^9$~GeV in the SM and $\simeq 10^{10}$~GeV in the MSSM, assuming that the Higgs field is minimally coupled to gravity. This decay channel vanishes for a coupling $\xi |H|^2 R$ with $\xi = \frac16$. However, in that case the inflaton can still decay to SM gauge bosons through the trace anomaly \cite{Gorbunov:2012ns}, which results in a reheating temperature of order $10^8$~GeV. 

In the supergravity formulation of the Starobinsky and related attractor models, additional SM chiral superfields can be added inside the logarithm of Eq.~(\ref{n-sK}). 
In that case, when one transforms to a canonical inflaton, its couplings with the SM fields appear as if they were coupled conformally to curvature with $\xi = \frac16$ \cite{eno9,Ema:2024sit}. Thus, barring a direct superpotential coupling between the inflaton and Standard Model fields, the decays of the inflaton are highly suppressed \cite{Endo:2006xg,egno4}. Nevertheless, we expect that in this case as well inflaton decay should proceed through 
the coupling of the K\"ahler potential to the trace anomaly \cite{Endo:2007sz}: $\mathcal{L} \supset (-K/6)T^\mu_\mu$. However for $\phi$-inflation (see Eq.~(\ref{phi})), $\langle \partial K/\partial\phi \rangle = 0$ and this contribution vanishes as well. Alternatively,
if the gauge kinetic function depends on the inflaton, decays to gauge bosons and gauginos may be the dominant source for reheating \cite{Endo:2006xg,egno4}. In what follows, we will not tie ourselves to any particular mode of decay, and simply treat the reheating temperature, $T_{\rm RH}$ as a (relatively) free quantity, subject only to general phenomenological constraints. 

We have already commented that we consider 
$\trh = 4$~MeV as a lower bound on the reheating temperature to ensure standard BBN \cite{tr4}. 
We nevertheless consider $\trh \ga 100$~GeV
as a reasonable lower bound to ensure the generation of a baryon asymmetry. Below the electroweak scale, sphaleron interactions are out-of-equilibrium and there are no interactions which are capable of shuffling any existing lepton or baryon asymmetries, thus requiring direct baryon number violation at scales below the electroweak scale. While we know of no proof that generating
a baryon asymmetry at lower energies is not possible, we consider the requirement of reheating above the electroweak scale as reasonably conservative.

For inflaton couplings approaching the non-perturbative limit, the decay rate, $\Gamma_\varphi
\sim H_{\rm I}$, where the Hubble parameter $H_{\rm I}$ during inflation is $H_{\rm I} = M/2 \sim \sqrt{\lambda} M_P/2$ for the Starobinsky model. From Eq.~(\ref{trh}),
this leads to $\trh \sim 2 \times 10^{15}$~GeV.
In supersymmetric models, high reheating temperatures generally lead to the overproduction of gravitinos which in turn lead to the overproduction of dark matter if the gravitino is not the lightest supersymmetric particle \cite{Moroi:1994rs,ego,Ellis:2015jpg}.

We can approximate the thermal production of gravitinos as ~\cite{ekn,enor,bbb,Pradler:2006qh,rst,Ellis:2015jpg,Ellis:2021kad,Eberl:2020fml,Eberl:2024pxr}
\begin{equation}
\label{gravitinoprodn}
Y_{3/2} \equiv \frac{n_{3/2}}{n_{\rm rad}} \simeq 0.003 \left(\frac{\Gamma_{\varphi}}{M_{p}}\right)^{1 / 2} \,,
\end{equation}
where $Y_{3/2}$ is the gravitino yield, $n_{\rm rad} = \zeta(3) T^3/\pi^2$.
If gravitinos decay after the lightest supersymmetric particle (LSP) freezes out, their decay products contribute to the LSP abundance
and we have the constraint
\beq
\Omega_{\mathrm{LSP}}h^2 = m_{\rm LSP} Y_{3/2} \left(\frac{n_\gamma}{2 \rho_c} \right)\leq 0.12 \, .
\eeq
Using Eq.~(\ref{trh}), we can translate this into a limit on the reheating temperature, $\trh \la 10^{10}$~GeV.

These considerations, combined with the CMB constraints from ACT DR6, significantly restrict the viable parameter space. In supersymmetric realizations, radiative corrections to the inflaton potential are suppressed by supersymmetry~\cite{ENOT,DreesXu,egkko}, protecting the flatness required for successful inflation.

\section{Results}
\label{sec:results}
We now present the constraints on attractor models of inflation we obtain using the {\it Planck} 2018~\cite{Planck}, BICEP/Keck 2018~\cite{BICEP2021}, ACT~DR6~\cite{ACT:2025fju}, and SPT-3G~\cite{SPT-3G:2025bzu} CMB data. Our analysis in the next subsection reveals significant tension between the canonical Starobinsky model (with $\alpha = 1$) and ACT DR6. There remains general agreement with the {\it Planck} 2018 data, but there is already some tension with the SPT-3G data. These tensions are reduced for $\alpha > 1$, and we show in subsequent subsections that the tensions are further relaxed in generalized attractor models with $k > 2$ as well as in deformed models with $\kappa < 1$. In all cases we base our analysis on the exact numerical values of the CMB observables, computed as discussed in Appendix~\ref{appC}.

\subsection{Standard Attractor Models}
\begin{figure*}[t!]
    \centering

\includegraphics[width=1\linewidth]{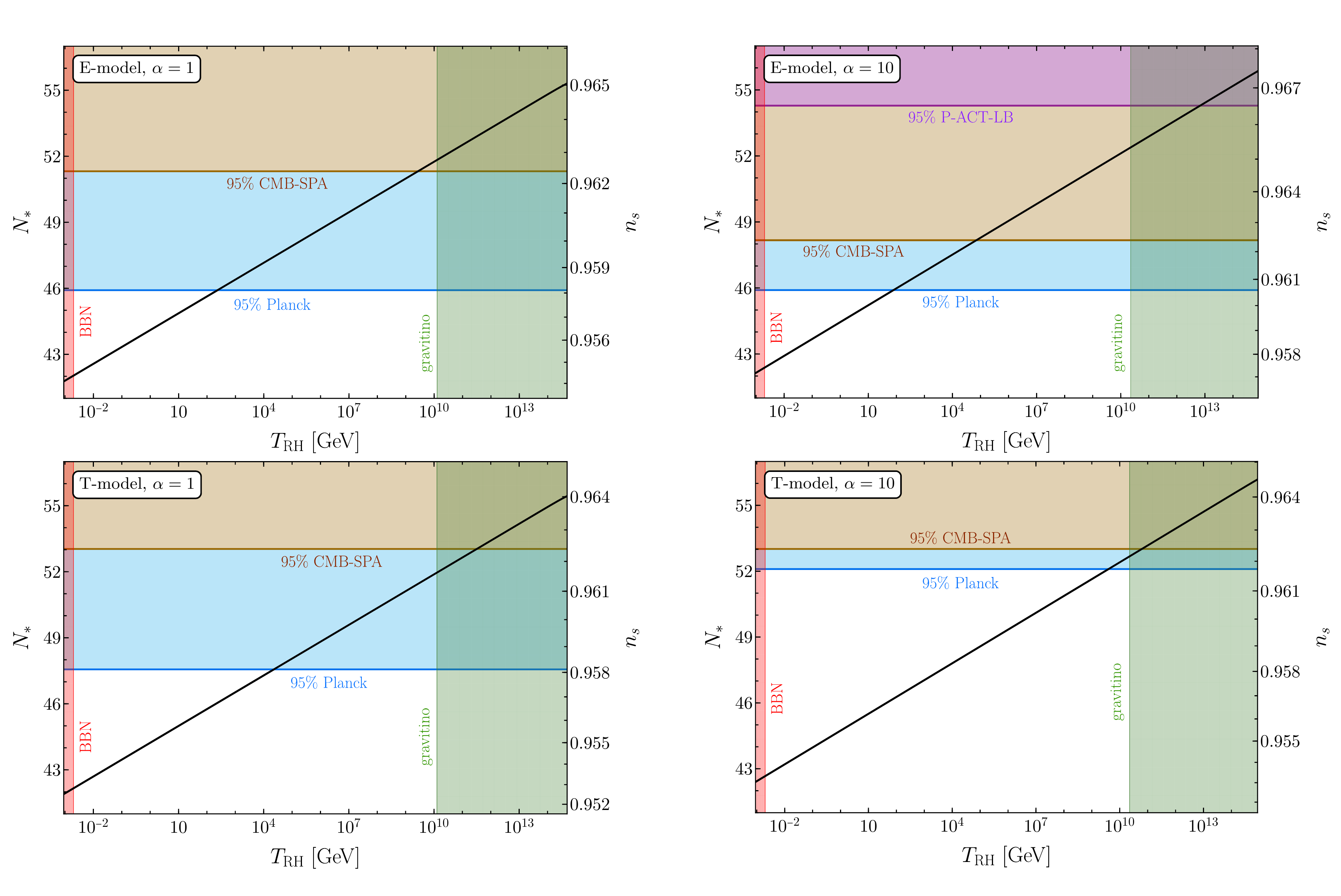}
    
\caption{The relation between $N_*$ (defined for a pivot scale of $0.05$~Mpc$^{-1}$) and the reheating temperature in the E-model (upper panels) and the T-model (lower panels), for $\alpha = 1$ (left panels) and $\alpha = 10$ (right panels). The horizontal shadings corresponds to the 95\% CL bounds from {\it Planck} 2018, CMB-SPA, and P-ACT-LB. The vertical shadings show the constraints of $T_{\rm RH}$ from Big Bang Nucleosynthesis (BBN) and gravitino production in supersymmetric models. Note that the limits on $n_s$ (right vertical axis) depend on $\alpha$, as the calculated value of $r$ depends on $\alpha$ as seen in Eq~(\ref{eq:gentmodr}).}
    \label{fig:NvsTfull}
\end{figure*}

\begin{figure*}[ht!]    
    \centering
    \begin{minipage}[t]{0.48\linewidth}
        \centering
        \includegraphics[width=\linewidth]{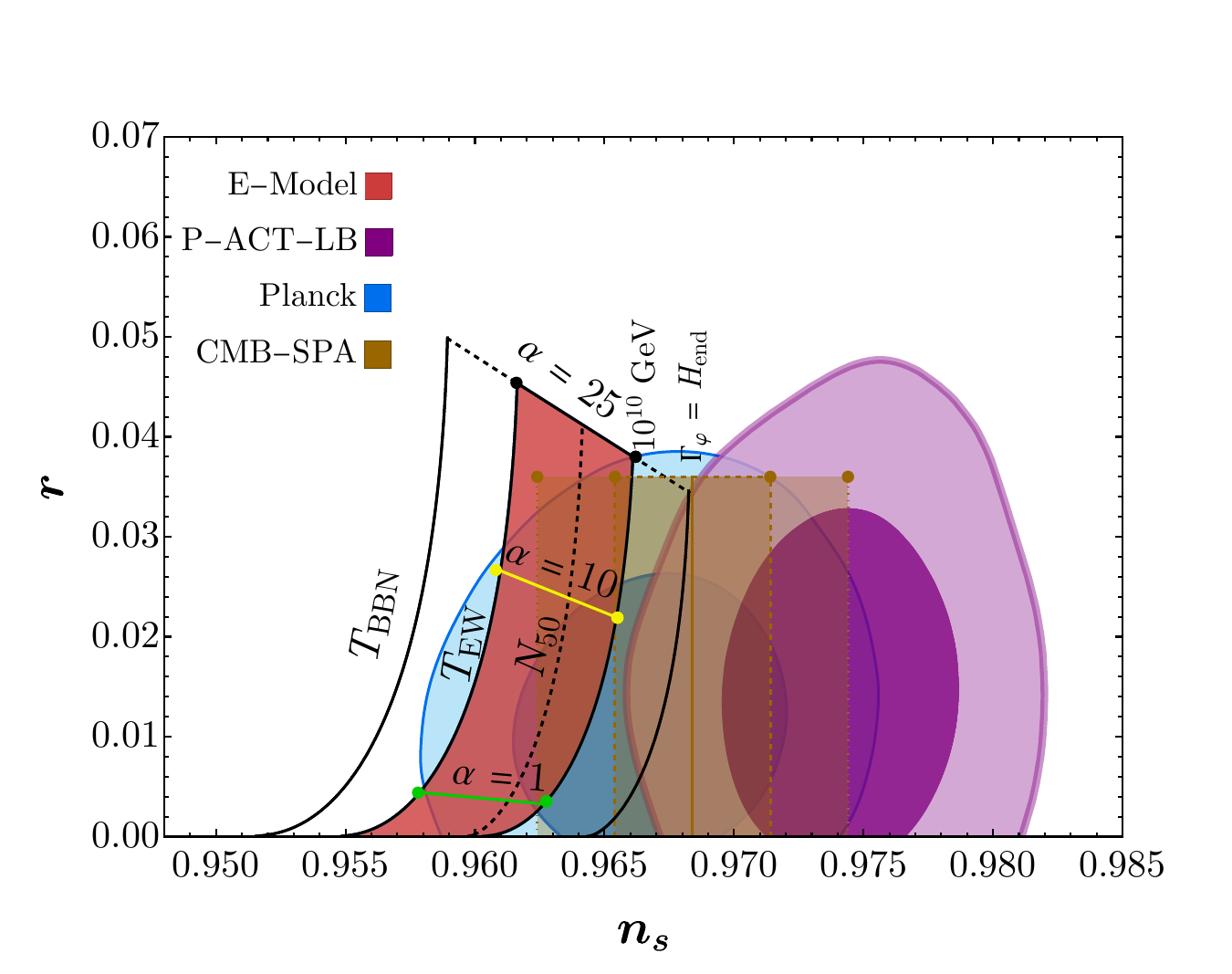}
    \end{minipage}
    \vspace{0.1cm} 
    \begin{minipage}[t]{0.48\linewidth}
        \centering
        \includegraphics[width=\linewidth]{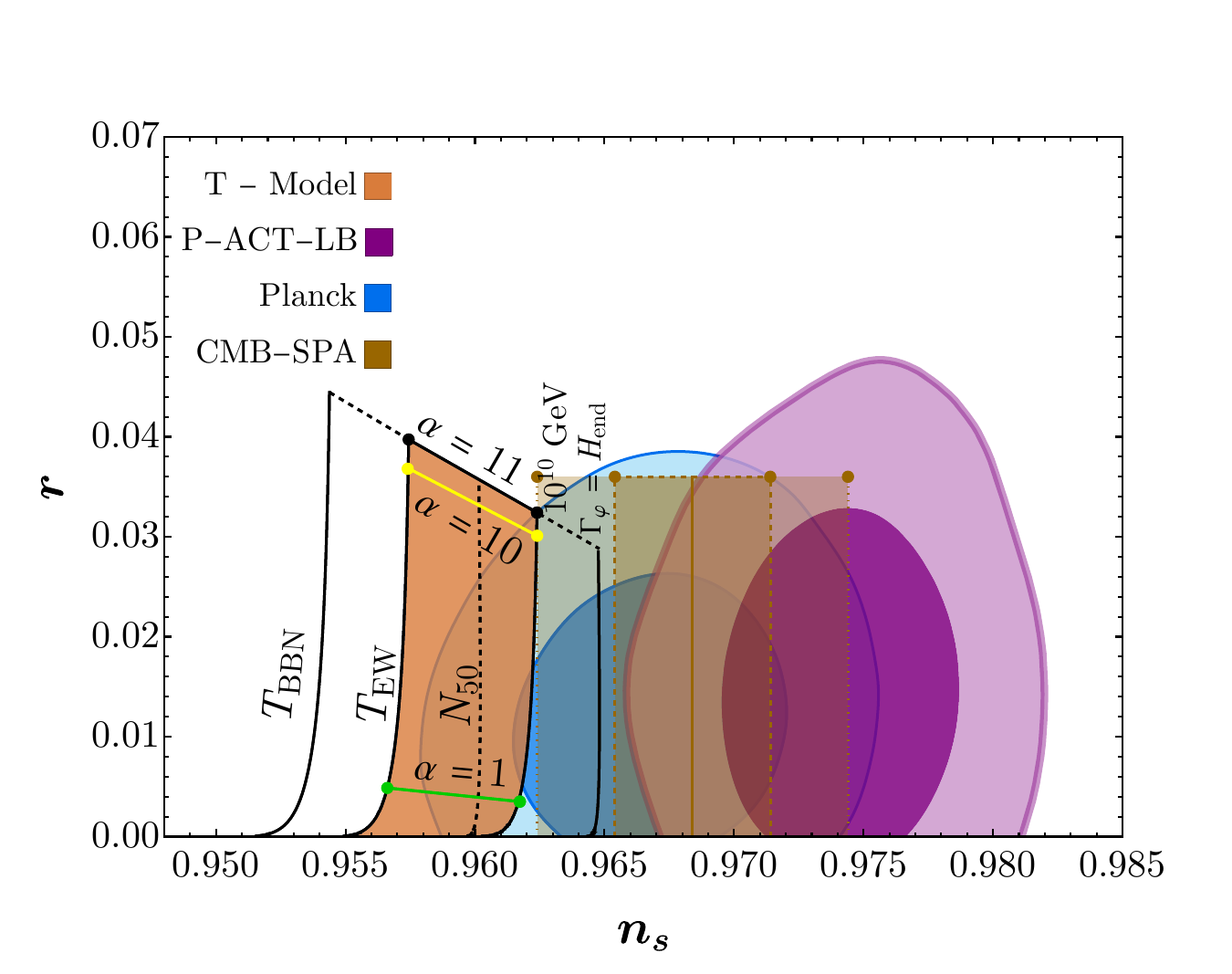}
    \end{minipage}
         \caption{Constraints on $\alpha$-attractor models showing the 68\% and 95\% CL contours from {\it Planck}/BICEP/Keck \cite{BICEP2021} (blue shadings), the P-ACT-LB combination \cite{ACT:2025tim} (purple contours) and the CMB-SPA dataset \cite{SPT-3G:2025bzu} (brown rectangles). Note that the latter provides only an $r$-independent limit on $n_s$. All pivot scales are taken as $k_*=0.05$ Mpc$^{-1}$ except for the {\em Planck} pivot scale for $r$, conventionally chosen to be $k_*=0.002$ Mpc$^{-1}$.  Left panel: $\alpha$-Starobinsky (E-model) predictions in the $(n_s, r)$ plane. Solid lines indicate reheating temperatures from $T_{\rm BBN}$ (4 MeV), $T_{\rm EW}$ (100 GeV), $10^{10}$~GeV (the gravitino bound), and $ 2 \times 10^{15}$~GeV (instantaneous reheating with $\Gamma_\varphi = H$). 
         The dashed line shows $N_* = 50$ for reference. 
         Right panel: As in the left panel for the T-model predictions.}
\label{fig:alpha1}        
\end{figure*}

As discussed in Section~\ref{sec:infdynamics},
the number of $e$-folds and $\varphi_*$ depend on the reheating temperature, which appears in Eq.~(\ref{eq:nstarreh}) through $\rho_{\rm rad}$. This in turn induces a dependence of $n_s$ on $\trh$. In Fig.~\ref{fig:NvsTfull}, we show this dependence for the E- and T-models (upper and lower panels) for $\alpha = 1$ and 10 (left and right panels). We note that although both $N_*$ (defined at the pivot scale of $0.05$ Mpc$^{-1}$) and $n_s$ depend on $\trh$, the relation between $N_*$ and $n_s$ has some model dependence.  The horizontal shading in Fig.~\ref{fig:NvsTfull} shows the 95\% CL lower limit on $n_s$ (blue for the limit from {\it Planck} 2018 alone, brown for the SPT-{\it Planck}-ACT combination (CMB-SPA), and purple for {\it Planck}-ACT-DESI combination (P-ACT-LB) seen only in the upper right panel for the E-model with $\alpha = 10$).
The vertical shading shows exclusions of reheating temperatures below 4 MeV from BBN on the left side of each panel, and on the right side temperatures above $\simeq 10^{10}$~GeV due to the gravitino bound discussed above. For $\alpha = 1$ there is a range of reheating temperatures consistent with the {\it Planck} 95\% lower bound on $n_s$ ($220 - 10^{10}$~GeV)
for the E-model and ($2\times 10^4 - 10^{10}$~GeV for the T-model. Neither model falls within the P-ACT-LB 95\% range for any possible reheating temperature.
The limits on $n_s$ depend on $r$ as can be seen in Fig.~\ref{fig:alpha1}, and  $r$ depends on $\alpha$ as seen in Eq.~(\ref{eq:gentmodr}). Thus limits on $n_s$ are model-dependent and are slightly different in each panel of Fig.~\ref{fig:NvsTfull}.
However, for $\alpha = 10$ in the E-model, $\trh \gtrsim 80$ or $10^5$~GeV are allowed by {\it Planck} 2018 and CMB-SPA respectively. 
The P-ACT-LB bound is now visible but requires 
$\trh \gtrsim 10^{13}$~GeV, in excess of the gravitino bound, but still less the temperature attained in instantaneous reheating. For the T-model with $\alpha = 10$, {\it Planck} is satisfied for $\trh \gtrsim 5 \times 10^9$~GeV, but CMB-SPA can not be satisfied simultaneously with the gravitino bound.

Fig.~\ref{fig:alpha1} displays the observational constraints on the $\alpha$-attractor models in the $(n_s, r)$ plane. We overlay the 68\% and 95\% CL contours from {\it Planck}/BICEP/Keck \cite{BICEP2021} (blue shadings), the P-ACT-LB combination \cite{ACT:2025tim} (purple contours) and the CMB-SPA dataset \cite{SPT-3G:2025bzu} (brown rectangles). Note that the {\it Planck} contours are provided at the WMAP pivot scale $k_*=0.002$ Mpc$^{-1}$, while the P-ACT-LB (and our predictions) correspond to $k_*=0.05$ Mpc$^{-1}$ for both $n_s$ and $r$. CMB-SPA does not provide an $r$-dependent limit on $n_s$.  The theoretical predictions show the trajectories for both the $\alpha$-Starobinsky (E-model) and T-model potentials given by Eqs.~(\ref{eq:emodel}) and~(\ref{eq:tmodel}), respectively. The left panel shows the E-model predictions for three  highlighted values of $\alpha = 1, 10$, and 25. The solid line trajectories correspond to specific reheating temperatures of 4 MeV (labeled BBN), 100 GeV (labeled $T_{\rm EW}$, $10^{10}$~GeV (as labeled) corresponding to the gravitino bound, and $2 \times 10^{15}$~GeV corresponding to instantaneous reheating (labeled as $\Gamma_\varphi = H$). The red shading highlights the range $10^2~{\rm GeV} < \trh < 10^{10}~{\rm GeV}$.  For reference, we also show the trajectory for $N_* = 50$ $e$-folds before the end of inflation. The right panel shows similar results for the T-model with $\alpha = 1, 10$, and 11 as examples. 

Results for the canonical Starobinsky model (E-model with $\alpha = 1$) are shown in the left panel of Fig.~\ref{fig:alpha1}. It predicts $n_s \simeq 0.961$ for $N_* = 50$, corresponding to a reheating temperature $\trh \simeq 5\times10^7$~GeV, which lies comfortably within the {\it Planck}+BICEP/Keck 95\% confidence region, but below the 95\% CL range of the CMB-SPA dataset, and falls significantly below the ACT DR6 preferred value of $n_s = 0.9752 \pm 0.0030$. The model remains observationally viable for $\alpha \lesssim 25$, beyond which it violates the upper limit  on $r$ for $\trh \le 10^{10}$~GeV. Efficient reheating (though in violation of the gravitino bound) maximizes $N_*$ and pushes $n_s$ toward higher values, improving agreement with ACT DR6 (purple shading). The right panel presents the corresponding constraints for T-models. These models predict systematically lower $n_s$ values than E-models models for the same $\alpha$, with the canonical case ($\alpha = 1$) yielding $n_s \simeq 0.960$ for $N_* = 50$ $e$-folds. This places T-models in even greater tension with ACT DR6 data. The observational constraints restrict T-models to $\alpha\lesssim 11$, a more stringent limit than in the $\alpha$-Starobinsky case. Both model classes require efficient reheating (high reheating temperatures) to approach the region preferred by the ACT~DR6 data.

\subsection{Generalized Attractor Models}\label{sec:phik}
\begin{figure*}[ht!]
    \centering
    \includegraphics[width=0.63\linewidth]{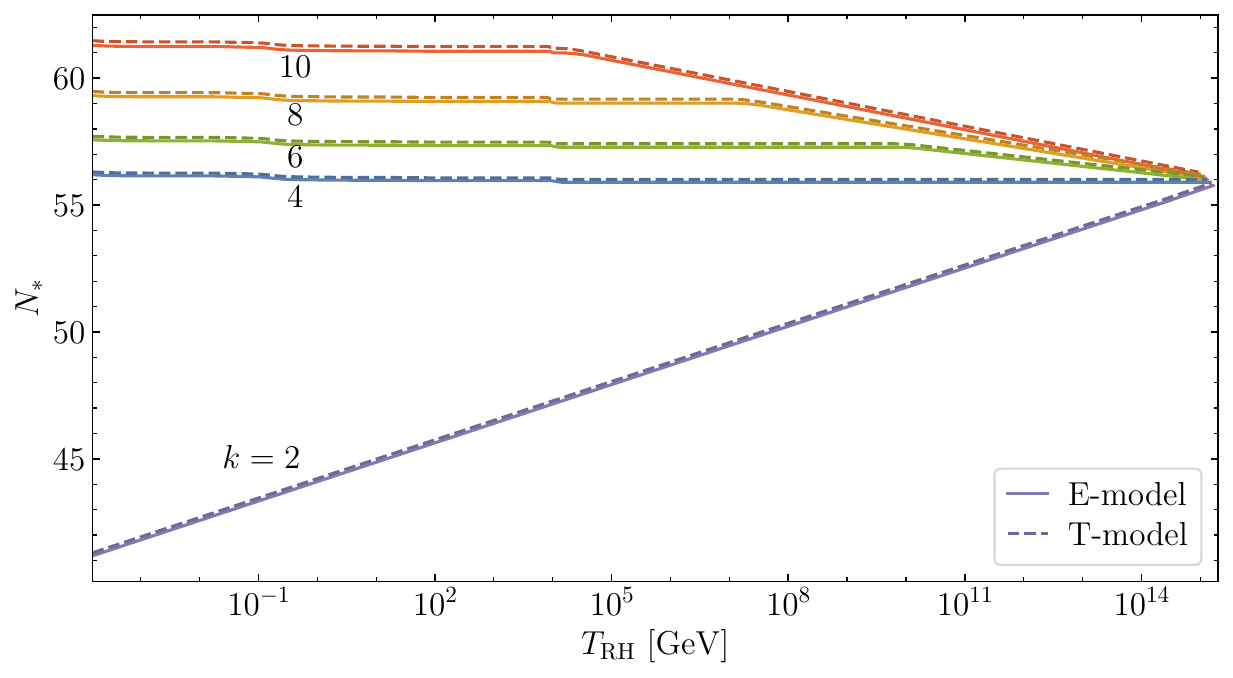}
    \caption{Plot of $N_*$ as a function of $\trh$ for different values of $k$ in the E- and T-models with $\alpha = 1$. }
    \label{fig:NvsTvsk}
\end{figure*}

We next examine generalized $\alpha$-attractor models with non-quadratic minima of the form $V \propto \varphi^k$, which modify both the inflationary dynamics and reheating phase. These models provide enhanced flexibility in reconciling theoretical predictions with the latest CMB observations, particularly the higher spectral index values favored by ACT DR6. The generalized E-model attractors are described by Eq.~\eqref{eq:alphastarogeneral} and the generalized T-models are described by Eq.~(\ref{eq:tmodelgeneral}).  The modification from the standard quadratic minimum ($k = 2$) to higher powers significantly alters the reheating dynamics through the effective equation of state $w_\varphi = (k-2)/(k+2)$, leading to enhanced $e$-fold numbers for fixed reheating temperatures. This effect is illustrated in Fig.~\ref{fig:NvsTvsk}, which shows the dependence of $N_*$ (defined at the pivot scale $k_* = 0.05$~Mpc$^{-1}$) for $k= 2, 4, 6, 8$, and 10. Solid lines correspond to the E-models and dashed lines correspond to the T-model, though the difference between the two is generally very small on the scale of this figure.  The lines for $k=2$ are identical to the ones given in Fig.~\ref{fig:NvsTfull} for $\alpha = 1$.
All of the lines converge when reheating is instantaneous, as the only effect from the shape of the potential is limited to the difference in the energy densities at the end of inflation, which is weakly dependent on $k$ (see Eqs.~(\ref{eq:phiendEk}) and (\ref{eq:phiendTk}) in Appendix~\ref{appB}). We note that $w_{\varphi}=1/3$ for $k=4$, and the contribution to $N_*$ from $\rho_{\rm rad}$ in Eq.~(\ref{eq:nstarreh}) drops out. Therefore, $N_* \simeq 56$ is mostly independent of $\trh$, except for the indirect dependence through $g_{\rm RH}$. 

For $k>4$, $w_{\varphi} > 1/3$, the sign of the contribution from $\rho_{\rm rad}$ in Eq.~(\ref{eq:nstarreh}) changes, and $N_*$ increases with decreasing $\trh$. However, in such a case, the self-interaction of the inflaton cannot be ignored, as it will drive the resonant growth of inflaton inhomogeneities, eventually fragmenting the inflaton condensate. The net effect is a transition to a radiation-dominated epoch, $w_{\varphi}\rightarrow 1/3$, where the dominant species are the nearly massless inflaton quanta~
\cite{Lozanov:2016hid,Lozanov:2017hjm,Garcia:2023eol,Garcia:2023dyf,Garcia:2024zir}. The time-scale for fragmentation is fixed for given values of $\lambda,k,\alpha$, and can be smaller than the reheating time-scale, when the now free inflaton quanta complete their decay into radiation. Read from right to left, Fig.~\ref{fig:NvsTvsk} follows the increasing trend of $N_*$ as a function of $\trh$ for $k>4$, until low reheating temperatures, for which reheating is completed after fragmentation, and the dependence of $N_*$ on $\trh$ becomes flat (see Eq.~(\ref{eq:Nstarka}) in Appendix~\ref{appB}).

\begin{figure*}[!ht]
    \centering
    \includegraphics[width=1\linewidth]{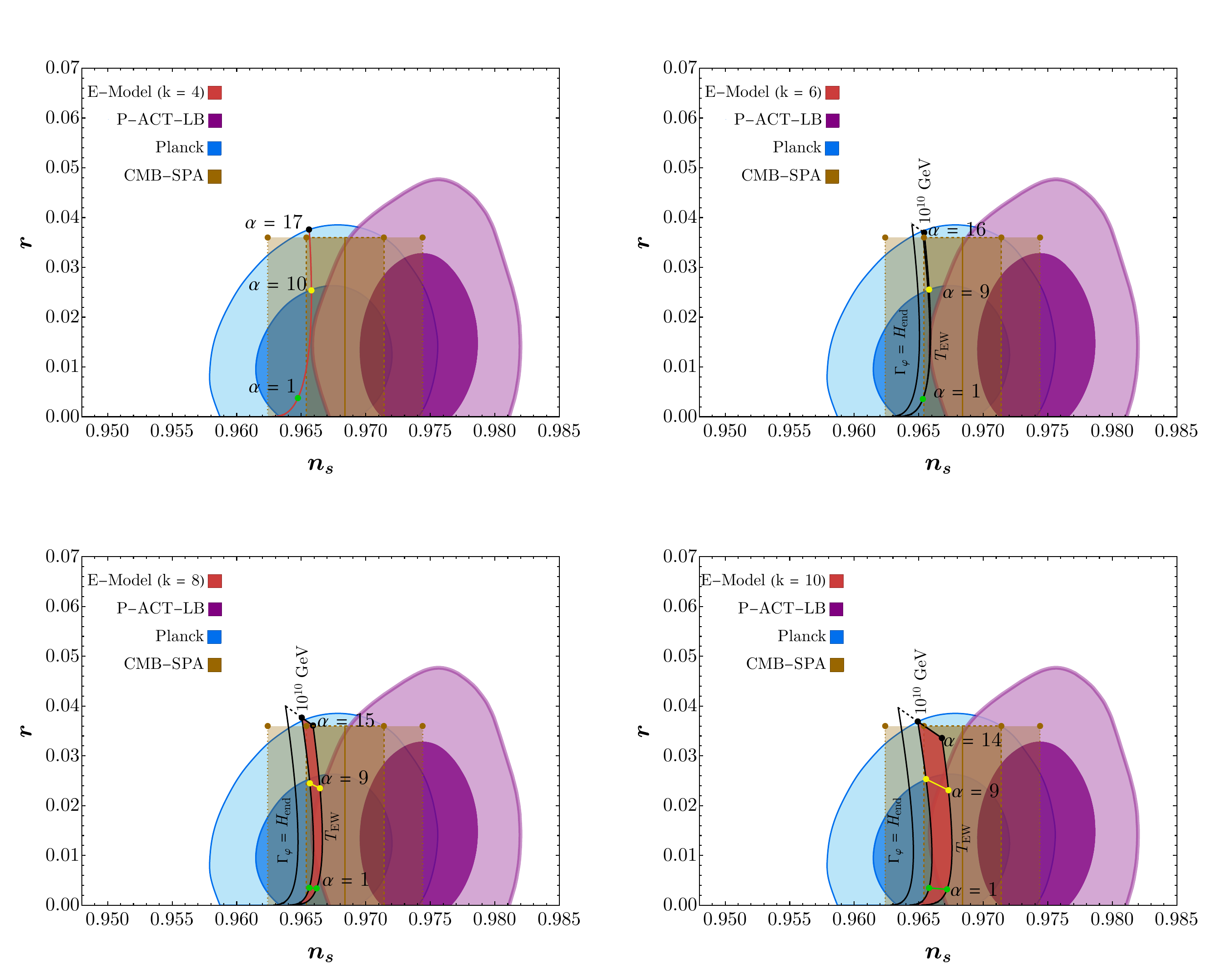}
    \caption{As in Fig.~\ref{fig:alpha1}, showing the constraints on generalized $\alpha$-Starobinsky (E-model) attractor models with $V \propto \varphi^k$ minima for $k = 4, 6, 8,$ and $10$. 
Shaded bands indicate the range of $N_*$ allowed by reheating temperatures from $T_{\rm EW}$ to $10^{10}$~GeV. The curve for $T_{\rm BBN}$ is degenerate with that shown for $T_{\rm EW}$ when fragmentation effects are included (see Fig.~\ref{fig:NvsTvsk}). For $k>4$, the curve for instantaneous reheating is to the left of the shaded band. Higher $k$ values systematically shift predictions toward larger $n_s$, improving consistency with the ACT~DR6 data. The constraint on $r$ limits all models to $\alpha \lesssim 14-17$.}
    \label{fig:fullstarok}
\end{figure*}

Fig.~\ref{fig:fullstarok} presents constraints on generalized E-model attractors given by Eq.~\eqref{eq:alphastarogeneral} for $k = 4, 6, 8,$ and $10$.
For $k = 4$ (upper left panel), the predictions track closely those of the standard E-model, with the observational constraints permitting $\alpha\lesssim 17$. Note that, in this case, there is no theoretical band, as there is no dependence of $n_s$ (and $r$) on $\trh$. Therefore we see only a single curve which varies with $\alpha$. The canonical case ($\alpha = 1$) predicts $n_s \simeq 0.965$ for the allowed $e$-fold range. This runs directly through the {\it Planck} 2018 ellipses and is also quite compatible with the CMB-SPA dataset. It is barely consistent with the 95\% bound from the P-ACT-LB dataset when $\alpha \simeq 5$. 

As $k$ increases to $6$ (upper right panel), 
there is only a narrow allowed range for  $N_*$ between $\simeq 56 - 58$. Note that in contrast to $k=2$, lower values of $\trh$ correspond to the right side of the trajectory strip. As for $k =4$, there is no shaded strip as the trajectories for  $\trh = 100$~GeV and $10^{10}$~GeV are nearly identical (as is the trajectory for $\trh = 4$~MeV) and these three trajectories in the upper right panel of Fig.~\ref{fig:fullstarok} are nearly degenerate. As can be seen in Fig.~\ref{fig:NvsTvsk}, the value of $N_*$ for instantaneous reheating is somewhat lower and yields the left trajectory  in that panel.   Concerning $n_s$ and $r$, we observe more substantial shifts: at $\alpha \simeq 5$ the prediction intersects the ACT DR6 95\% confidence region at $N_* \simeq 58$, corresponding to $n_s \simeq 0.966$. The canonical model ($\alpha = 1$) yields $n_s \simeq  0.965$, representing a notable enhancement compared to the standard ($k=2$) case. The upper limit on $\alpha$ is slightly lowered to $\alpha < 16$. The trend continues for $k = 8$ and $10$ (lower panels), where the predictions systematically shift toward higher $n_s$ values. For these values of $k$, there is a slight difference in $N_*$ for $\trh = T_{\rm EW}$ and $10^{10}$~GeV, resulting in the thin shaded strip seen in the lower two panels. The trajectory for $\trh = 4$~MeV remains degenerate with the one for $\trh = T_{\rm EW}$. For $k = 10$ with $\alpha \simeq 5$, the model predictions overlap substantially with both the ACT DR6 and CMB-SPA preferred regions. The maximum viable $\alpha$ remains approximately constant at $\alpha \sim 16$ for all $k$ values, determined by the tensor-to-scalar ratio constraint $r \lesssim 0.04$. 

\begin{figure*}[t!]
    \centering

\includegraphics[width=1\linewidth]{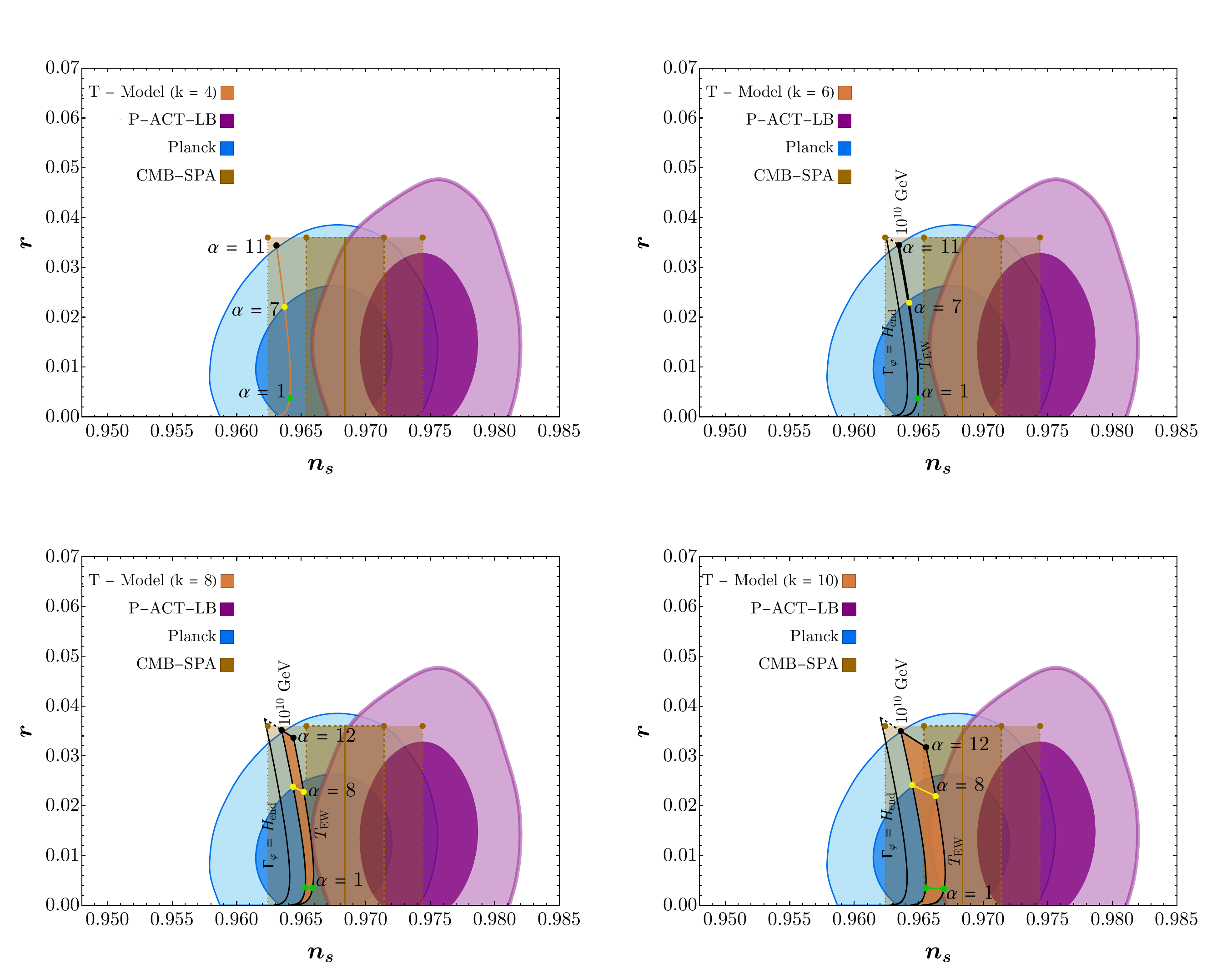}
\caption{As in Fig.~\ref{fig:fullstarok}, showing the constraints on generalized T-model attractors with $V \propto \varphi^k$ minima for $k = 4, 6, 8,$ and $10$.
While T-models predict systematically lower $n_s$ than E-models, increasing $k$ substantially improves compatibility with observations. For $k = 10$, even the canonical model with $\alpha = 1$ approaches the observational confidence regions.}
    \label{fig:fulltplots}
\end{figure*}

Fig.~\ref{fig:fulltplots} displays the corresponding analysis for the generalized T-model attractors given by Eq.~\eqref{eq:tmodelgeneral}. While T-models inherently predict lower spectral indices than their E-model counterparts, the higher-$k$ variants demonstrate significantly improved agreement with observations. For $k = 4$ (upper left), $\alpha \lesssim 11$ yields predictions that  are fully consistent with {\it Planck} 2018, within the 95\% CL of CMB-SPA, but
remain below the ACT DR6 values.
However, increasing to $k = 6$ (upper right) allows $N_*$ to increase and shifts the trajectory closer to the P-ACT-LB observational contours, with $n_s$ reaching $\sim 0.965$ for $\alpha = 1$ at $N_*\simeq 58$ $e$-folds. As for the E-model, the trajectories for $\trh = 10^{10}$~GeV, 100 GeV, and 4 MeV are nearly degenerate. The most dramatic improvement occurs for $k = 8$ and $10$ (lower panels). For $k = 10$ with $\alpha \simeq 1$, the model predictions enter the ACT DR6 95\% confidence region, reaching $n_s \simeq 0.965-0.967$ over the allowed $e$-fold range of $N_* \simeq 59 -61.5$ for $\trh = 100 - 10^{10}$~GeV.

\subsection{Deformed No-Scale Attractors}
\label{sec:def}

\begin{figure*}[t!]
    \centering
    \begin{minipage}[t]{0.48\linewidth}
        \centering
        \includegraphics[width=\linewidth]{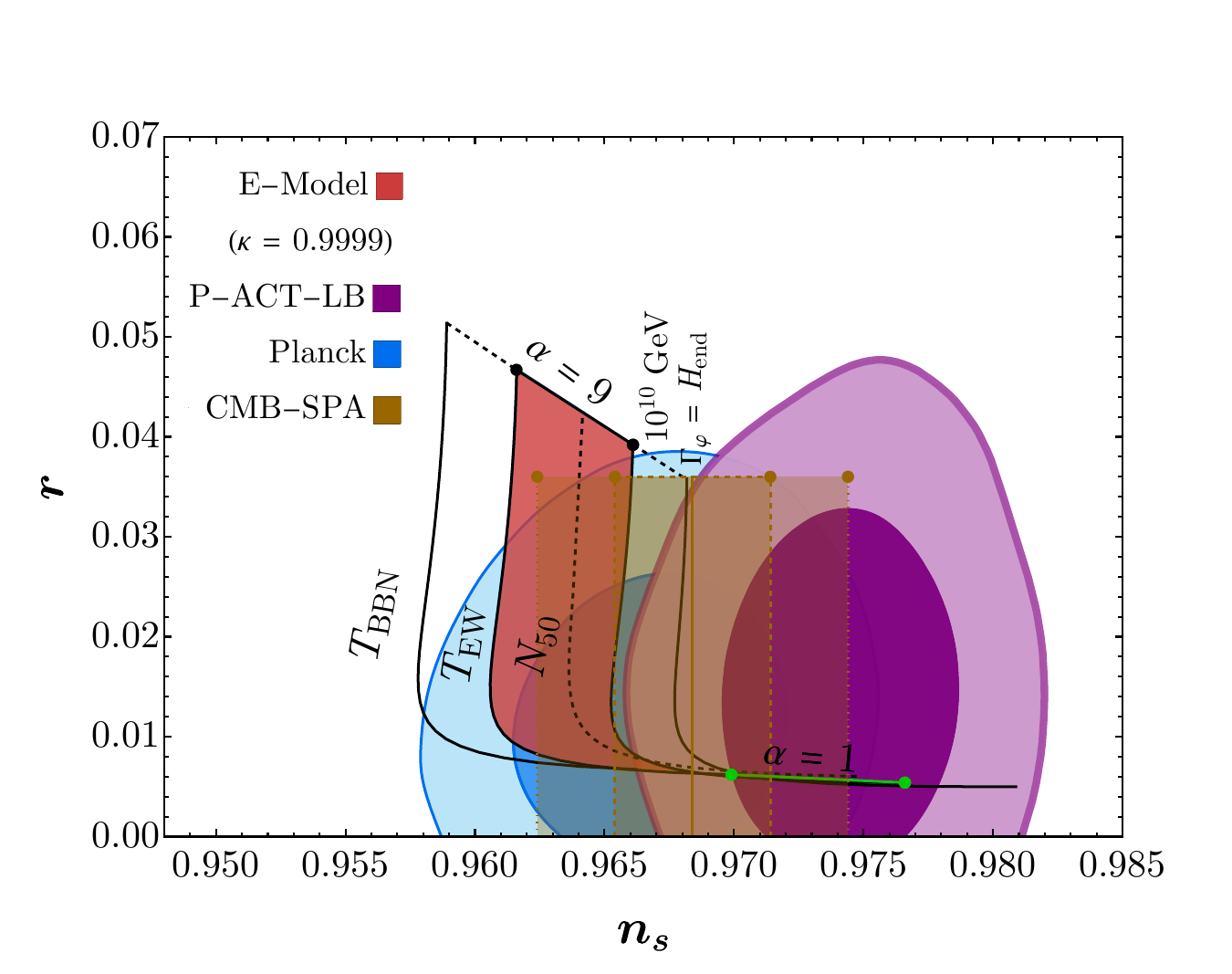}
    \end{minipage}
    \hfill
    \begin{minipage}[t]{0.48\linewidth}
        \centering
        \includegraphics[width=\linewidth]{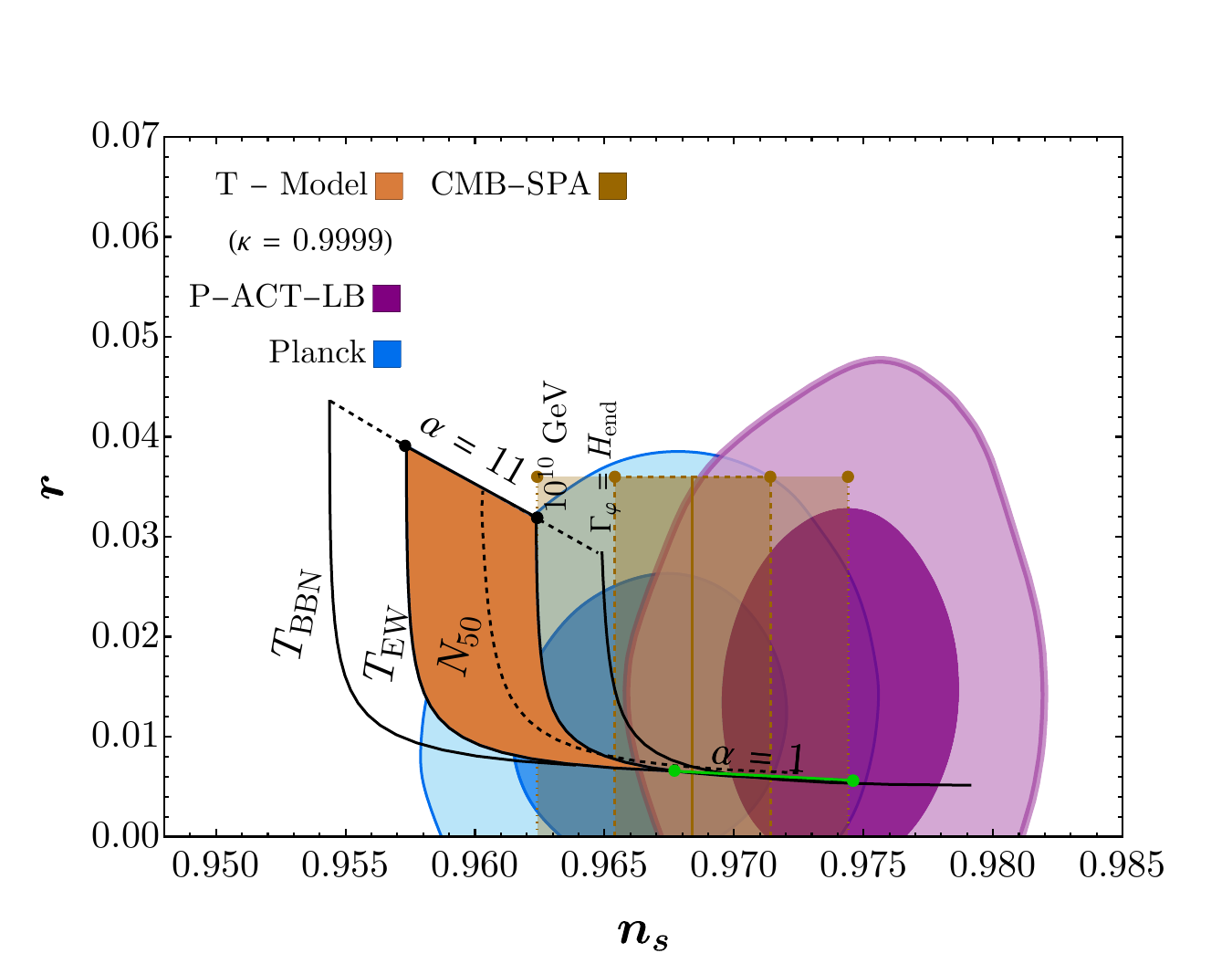}
    \end{minipage}    
        \caption{As in Fig.~\ref{fig:alpha1}, showing the constraints on deformed E-model attractors (\ref{eq:modstaro}) (left panel) and T-model attractors (\ref{eq:modtmodel}) (right panel) with the deformation parameter $\kappa = 0.9999$.  Observational contours and 
        theoretical trajectories for various $\alpha$ values demonstrate how minimal modifications ($|1-\kappa| \sim 10^{-4}$) shift predictions through the centers of observational confidence regions. The canonical model ($\alpha = 1$) with $\kappa = 0.9999$ achieves remarkable agreement with the P-ACT-LB dataset while maintaining consistency with the $r$ constraint. Red-shaded band (E-model) and orange-shaded band (T-model) indicate reheating uncertainty from $T_{\rm EW}$ to $10^{10}$~GeV. }        
        \label{fig:alphastarokappa09998}
\end{figure*}

We examine next deformed $\alpha$-attractor models, introducing a parameter $\kappa < 1$ that controls this deviation as in Eq.~(\ref{eq:modstaro}) for the E-model and in Eq.~(\ref{eq:modtmodel}) for the T-model. These modifications preserve the attractor nature of the models while providing enhanced flexibility to mitigate the tensions with different CMB datasets. Fig.~\ref{fig:alphastarokappa09998} presents the modified E-model (left) and T-model (right) predictions for $\kappa = 0.9999$. This minimal deviation from the standard attractor ($\kappa = 1$) creates a distinctive trajectory that passes directly through the centers of the observational confidence regions. For the E-model and the canonical case ($\alpha = 1$) with $\kappa = 0.9999$, the model predicts $n_s \simeq 0.968-0.975$ and $r \simeq 0.007-0.005$ across the reheating range from $T_{\rm EW}$ to $10^{10}$~GeV.  This represents a substantial shift from the standard Starobinsky prediction. The value of $n_s$ is now {\em high} compared with both {\it Planck} 2018 and CMB-SPA but is now in agreement with P-ACT-LB preferred range.  The modification preserves the fundamental attractor structure, evidenced by the convergence of different $\alpha$ trajectories at large $N_*$, while introducing sufficient flexibility to reconcile diverse observational constraints. As discussed in the previous Section and Appendix~\ref{appA}, such modifications arise naturally within the no-scale supergravity framework through appropriate choices of the K\"ahler potential and superpotential. 

Similar results are found for the deformed T-model, as shown in the right panel of Fig.~\ref{fig:alphastarokappa09998}. For the same value of $\kappa = 0.9999$ and $\alpha = 1$, the shift in $n_s$ is less extreme giving $n_s \simeq 0.970 - 0.977$ for the same range in reheating temperatures of $T_{\rm EW}$ to $10^{10}$~GeV, which is, within uncertainties, in agreement with all three datasets considered. 
Similar effects on $n_s$ in deformed models were seen in \cite{deform}. Note that for $\alpha > 1$,
the effect of the deformation is minimal and the range in $n_s$ is similar to that found in the undeformed attractor models. We now find upper limits of $\alpha \le 9$ (E-model) and $\alpha \le 11$ (T-model). 

\begin{figure*}[ht!]
    \centering
    \includegraphics[width=1\linewidth]{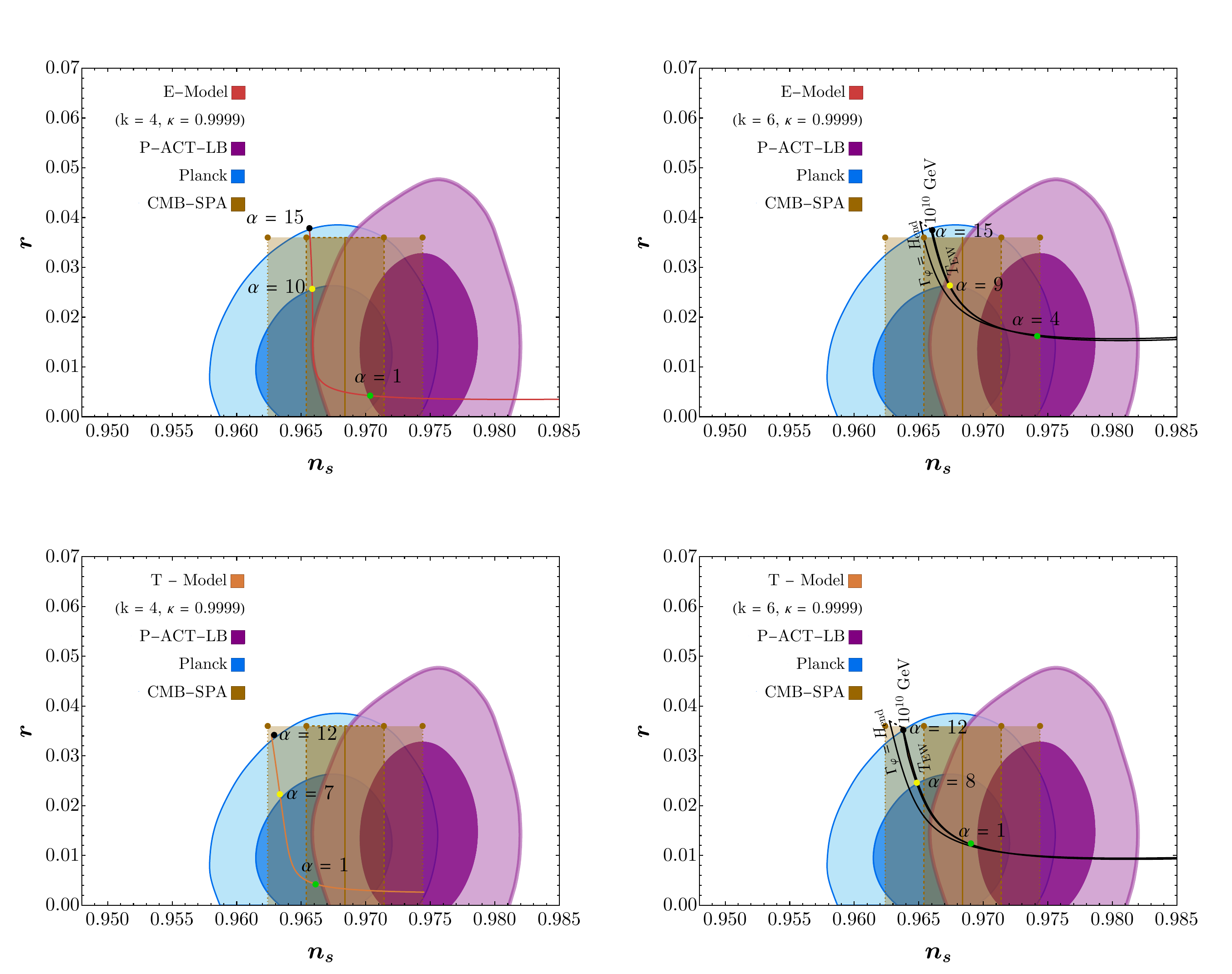}
        \caption{As in Fig.~\ref{fig:alpha1}, showing the constraints on deformed attractor models with $\kappa = 0.9999$ for generalized potentials with $k = 4$ (left panels) and $k = 6$ (right panels). Both E-models (upper panels) and T-models (lower panels) show significant shift in the predictions for $n_s$ for $\alpha \sim 1$. These results demonstrate that deformations of order $\kappa \sim 10^{-4}$ can fully reconcile attractor models with current observations.}
        \label{fig:kappa0p99999}
\end{figure*}

Fig.~\ref{fig:kappa0p99999} extends this analysis to include generalized deformed attractors with non-quadratic minima. The top panels display modified E-models with $\kappa = 0.9999$ for $k = 4$ (left) and $k = 6$ (right). For $k = 4$ with $\alpha = 1$, the model predicts $n_s \simeq 0.970$ and $r \simeq 0.004$. These models achieve exceptional agreement with the {\it Planck} 2018, CMB-SPA, and P-ACT-LB datasets, demonstrating that perturbations of order $10^{-4}$ to the attractor structure can fully reconcile present observational tensions. 
For $k=6$, the $\alpha = 1$ prediction is high relative to {\it Planck} 2018 and CMB-SPA, but agreement with all three datasets is achieved for slightly higher $\alpha \le 4$.  Indeed, $k = 6$ yields the prediction $n_s \simeq 0.974$ with $r \simeq 0.016$ for $\alpha =4$ and $N_* = 56-58$ $e$-folds. The upper limits on $\alpha$ from {\it Planck} 2018 are $\alpha \le 15$ for $k=4$ and $\alpha \le 16$ for $k=6$.

The bottom panels of Fig.~\ref{fig:kappa0p99999}  present the corresponding modified T-model predictions.  For the same value of $\kappa$, the shift in $n_s$ is smaller for the T-models. For $k = 4$ with $\kappa = 0.9999$ and $\alpha = 1$, we find $n_s \simeq 0.966$ and $r \simeq 0.004$, which satisfies the CMB-SPA constraint but remains marginally outside the P-ACT-PB 95\% confidence region. However, for $k = 6$ case the shift in $n_s$ is more substantial, leading to better concordance with the observations. In this case,  for $\alpha = 1$, the predictions reach $n_s \simeq 0.969$ with $r \simeq 0.012$, entering the overlap region between all observational constraints.

To summarize our findings, we identify three exemplary scenarios for reconciling $\alpha$-attractor models with current observations. (a) Standard attractors ($\kappa = 1$, $k = 2$) remain viable with $\alpha \lesssim 40$ for E-models and $\alpha \lesssim 11$ for T-models when requiring efficient reheating. These models satisfy the combined CMB-SPA constraints but show tension with ACT DR6 observations. (b) Generalized attractors ($\kappa = 1$, $k \geq 8$) naturally accommodate higher $n_s$ values through modified reheating dynamics, achieving compatibility with all datasets for moderate $\alpha$ values. (c) Deformed attractors ($\kappa \simeq 0.9999$) provide the most complete reconciliation, with even canonical models ($\alpha = 1$) capable of matching the full range of observational preferences. The deformed and generalized variants provide theoretical frameworks that can accommodate the full spectrum of current observations while maintaining the appealing features of attractor models of inflation.

\section{Conclusions}
\label{sec:conclusions}

Currently, models of inflation are constrained by CMB observations that determine the spectral tilt of scalar perturbations, $n_s$, and bound the tensor-to-scalar ratio, $r$. The former indicates a definite departure from purely scale-free perturbations, as is expected in many models of single-field inflation. The convergence of a value of $n_s$ near 0.97 began with WMAP~\cite{wmap}, and {\it Planck} 2018 data~\cite{Planck} reached a precision exceeding 0.5 \%. Among the well studied and relatively simple single-field models of inflation, the Starobinsky model \cite{Staro} appeared to be in best agreement with {\it Planck} 2018 data. For example, for $N_* \sim 52$ (corresponding to a reheating temperature $\trh \sim 10^{10}$~GeV), this model predicts $n_s \sim 0.963$, to be compared with the 68\% CL range of 0.961-0.969 determined by {\it Planck}. Similar agreement is attained in the related T-models~\cite{Carrasco:2015rva},
which predict a slightly lower value for $n_s$ for the same reheating temperature. 

These models are special cases of a more general class of model known as $\alpha$-attractors. The Starobinsky model (and its T-model cousin) can be formulated in the context of no-scale supergravity \cite{ENO6,Cecotti,GKMO} with a K\"ahler potential given by Eq.~(\ref{n-sK}) with $\alpha = 1$. The K\"ahler potential determines the metric in field space, and the no-scale potential characterizes a field space of constant curvature ($R = 2/3$). As is evident from the form of Eq.~(\ref{n-sK}), choosing $\alpha \ne 1$ yields field spaces of arbitrary constant curvature $R = 2/3\alpha$, generating the class of $\alpha$-attractor models~\cite{ENO7,Kallosh:2013hoa,T-model,KLR}. Typically, models with $\alpha > 1$ predict similar (though slightly higher) values of $n_s$
but significantly higher values of $r$. For example, the Starobinsky model with $N_* = 52$
predicts $r \simeq 0.0039$, to be compared with the {\it Planck}/BICEP/Keck upper limit of $r < 0.036$ \cite{BICEP2021}, whereas this limit is saturated for $\alpha = 25$ in the E-model, and $\alpha = 11$ in the T-model.

In addition to the 1-parameter class of $\alpha$-attractors generated by the prefactor of the K\"ahler potential in Eq.~(\ref{n-sK}),
we have considered generalized versions of the E- and T-models as described by Eqs.~(\ref{eq:alphastarogeneral}) and (\ref{eq:tmodelgeneral}), respectively. These models can also be also formulated within no-scale supergravity by altering the superpotential as described in Appendix~\ref{appA}. 
These models have the property that the potential 
takes for the form $V(\varphi)\sim \varphi^k$, when expanded about the origin. 
The $k=4$ models have the interesting feature that $N_*$ does not depend on the reheating temperature and therefore the predictions of $n_s$ and $r$ depend only on $\alpha$ as shown in the upper left panels of Figs.~\ref{fig:fullstarok} and \ref{fig:fulltplots}.

The generalized models with $k > 2$, 
typically provide larger values of $n_s$. For example, the E-model with $k=6$ predicts a narrow 
range in $n_s \sim 0.965$ in excellent agreement with {\it Planck} 2018 and the CMB-SPA data combination, though it still falls slightly below the 95\% CL region of the P-ACT-LB combination. 
The apparent preference for $k >2$ of certain datasets might indicate modified reheating dynamics, since higher $k$ values lead to stiffer equations of state during inflaton oscillations, naturally extending the number of $e$-folds and raising $n_s$. 
  
As discussed in \cite{accident,accid}, many models of inflation have an accidental nature to them, as the form of the potential may rely on a careful adjustment of some coupling to realize successful inflation. For example, the ratio of the quadratic and cubic couplings in Eq.~(\ref{wz}) must be extremely close to $2/3\sqrt{3}$ to obtain the Starobinsky potential. Slight deformations may lead to very different predictions to $n_s$ and $r$ \cite{ENO6}. Hence we have also studied a further generalization of the E- and T-models by allowing for small deformations of the potential as given by Eqs.~(\ref{eq:modstaro}) and (\ref{eq:modtmodel}),  characterized by a parameter $\kappa \simeq 1$. These too can be derived from no-scale supergravity as discussed in more detail in Appendix \ref{appA}. 
The required fine-tuning $|\kappa - 1| \sim 10^{-4}$ suggests that if such modifications exist in nature, they must arise from highly-suppressed corrections, possibly from Planck-scale physics, string-theory effects, or radiative corrections.

We have found that the deformed E- and T-models can lead to significant changes in $n_s$, particularly for $\alpha \simeq 1$. For example, 
for $\kappa = 0.9999$, $k=2$, and $\alpha = 1$, the E-model predicts $n_s \simeq 0.968-0.975$ with $r \simeq 0.007 - 0.005$ for a reheating temperature between 100 and $\sim 10^{10}$~GeV. For $\alpha > 1$, the predictions approach those of the undeformed models. For $k > 2$, similar results are found, though the range of $n_s$ is smaller due to the lack of sensitivity to the reheating temperature. Similar effects were found in the presence of deformations caused by combining inflation with a Grand Unified Theory such as SU(5) or SO(10) \cite{deform}.

Next-generation experiments will provide definitive tests of the $\alpha$-attractor paradigm. LiteBIRD will provide full-sky B-mode polarization maps with sensitivity $\sigma(r) \simeq 0.001$, enabling direct constraints on $\alpha$ and discrimination between E-models and T-models via their different $(r, n_s)$ predictions~\cite{Hazumi:2019lys}. If future measurements converge on $n_s \simeq 0.965$ with $r < 0.001$, this would strongly favor small-$\alpha$ E-models, pointing toward specific supergravity embeddings. Conversely, detection of $r \sim 0.01$ would indicate $\alpha \sim 1$-$10$.

Our analysis has shown how attractor models of inflation can navigate successfully the challenges posed by increasingly precise observations of the CMB. While individual experiments show preferences for different parameter regions, the theoretical flexibility of the framework, through generalized potentials, modified attractors, and varying reheating scenarios, ensures its continued viability. 
As we approach the era of next-generation CMB experiments, attractor models remain among the most compelling and testable paradigms for cosmological inflation, with the potential to reveal fundamental aspects of quantum gravity and early universe physics.

\subsection*{Acknowledgments}
The work of J.E. was supported by the United Kingdom STFC Grant ST/T000759/1. The work of M.A.G.G. was supported by the DGAPA-PAPIIT grant IA100525 at UNAM, and the SECIHTI ``Ciencia de Frontera” grant CF-2023-I-17. The work of K.A.O. was supported in part by DOE grant DE-SC0011842 at the University of Minnesota. The work of S.V. was supported by the Kavli Institute for Cosmological Physics at the University of Chicago.

\section*{Appendices}
\appendix

\renewcommand{\thesubsection}{\Alph{subsection}}
\setcounter{equation}{0}
\renewcommand{\theequation}{\thesubsection.\arabic{equation}}
\subsection{Generalized Inflationary Attractors in No-Scale Supergravity}
\label{appA}
\setcounter{equation}{0}

The generalized $\alpha$-attractor potentials introduced in Eqs.~(\ref{eq:alphastarogeneral}) and~(\ref{eq:tmodelgeneral}) are not just phenomenological constructions, but can be derived rather easily from fundamental supergravity theories. This connection provides both theoretical motivation and constraints on the allowed parameter space. We demonstrate here how these potentials emerge from no-scale supergravity \cite{no-scale,LN} with specific choices of the superpotential.

We start with the no-scale K\"ahler potential given in Eq.~(\ref{n-sK}).
As discussed earlier, the parameter $\alpha$ determines the curvature of the K\"ahler manifold, with $R = 2/(3\alpha)$ \cite{EKN1}, and recall that the scalar potential in supergravity is given by:
\begin{equation}
V = e^{K/M_P^2}\left[K^{i\bar{j}}D_iW D_{\bar{j}}\overline{W} - \frac{3}{M_P^2}|W|^2\right] \,,
\end{equation}
where $D_iW = \partial_iW + \frac{K_i}{M_P^2} W$ is the K\"ahler covariant derivative, and $K^{i\bar{j}}$ is the inverse K\"ahler metric.

In Section \ref{sec:no-scale}, we described the simple derivation of the Starobinsky potential (which is identical with that in the E-model with $\alpha = 1$) from either of the superpotentials (\ref{wz}) or (\ref{W3}), which both lead to Eq.~(\ref{starpot}) once $T$ or $\phi$, respectively  is fixed (stabilized) \cite{ENO6,ENO7}. Similarly, the T-model potential (\ref{eq:tmodel}) with $\alpha = 1$ can be derived using~\cite{GKMO}
\beq
W = \frac14\sqrt{\lambda} \left(M_P \phi^2 - \frac16 \frac{\phi^4}{M_P} \right) \, ,
\label{Wtmod}
\eeq
with $T = \frac{1}{2}M_P$ fixed, or 
\beq
W = \sqrt{\frac{3}{4}} \sqrt{\lambda}~M_P~\phi~(2T) \left( \frac{2T/M_P-1}{2T/M_P+1} \right)\, ,
\eeq
with $\phi = 0$ fixed. 

The generalization of the T-model potentials with  $V(\varphi) \propto \varphi^k$  near the minimum but still with $\alpha = 1$ are also easily expressed in terms of $\phi$ and $T$ 
\beq
\begin{aligned}
W = 3^{\frac12-\frac{k}{4}}\sqrt{\lambda} M_P^3\left(\frac{1}{k+2} \left(\frac{\phi}{M_P}\right)^{\frac{k}{2}+1} \right. \\
\left. - \frac{1}{3(k+6)} \left( \frac{\phi}{M_P}\right)^{\frac{k}{2}+ 3} \right) \, .
\label{Wtmodk}
\end{aligned}
\eeq
or 
\beq
W = \sqrt{\frac{3}{4}} \sqrt{\lambda}~M_P~\phi~(2T) \left( \frac{2T/M_P-1}{2T/M_P+1} \right)^{\frac{k}{2}} \, ,
\label{WTk}
\eeq
The superpotential for the generalized E-model cannot be  
expressed as a simple polynomial of $\phi$ as in Eq.~(\ref{wz}) for $k=2$ or as in Eq.~(\ref{Wtmodk}) for the generalized T-models, but can be expressed in terms of a hypergeometric function ${{}_2}F_1(-1 + \frac{k}{2}, 1 + \frac{k}{2}, 2 + \frac{k}{2}, -\frac{\phi}{\sqrt{3}})$.

There are several general forms of the superpotential that lead to relatively simple forms for the scalar potential.  These are useful for finding the corresponding superpotential for many of the types of models discussed here. For example, taking
\beq
W= A \sqrt{\lambda} M_P^3 \left(\frac{T}{M_P}-\frac12\right) f\left(\frac{\phi}{M_P} \right) \, ,
\label{Wfphi}
\eeq
gives 
\beq
V = 3 A^2 \lambda M_P^4 \frac{|f(\frac{\phi}{M_P})|^2}{(3-|\phi/M_P|^2)^2} \, ,
\eeq
when one fixes $T = \frac{1}{2} M_P$, and the canonical inflaton is given by Eq.~(\ref{phi}). 
Similarly, 
\beq
W= A \sqrt{\lambda} M_P^2 \phi \left(\frac{2T}{M_P}\right) f\left(\frac{T}{M_P}\right) \, ,
\label{simpT}
\eeq
gives 
\beq
V = 4 A^2 \lambda M_P^4 \frac{|T|^2}{(T+{\bar T})^2} \bigg|f\left(\frac{T}{M_P}\right) \bigg|^2 \, ,
\eeq
when one fixes $\phi = 0$ and the canonical inflaton is given by Eq.~(\ref{canT}). 

In this case, the generalized E-models are very simply obtained from Eq.~(\ref{Wfphi}) with
\beq
f\left(\frac{\phi}{M_P}\right) = \left(\frac{\phi/M_P}{\phi/M_P+\sqrt{3}}\right)^{\frac{k}{2}} \left(3-\phi^2/M_P^2\right) \, ,
\eeq
using $A^2 = 2^{k-2}$. We then fix the fields as
\begin{equation}
\langle T \rangle = \frac{1}{2}M_P, \quad \phi = \bar{\phi} \quad \text{(real modulus)} \,,
\end{equation}
and use the canonical field normalization 
in Eq.~(\ref{phi}).
Alternatively, the same potential can be obtained from Eq.~(\ref{simpT}), and 
\beq
f\left(\frac{T}{M_P} \right) = \left(1-\frac{M_P}{2T}\right)^\frac{k}{2} \, ,
\eeq
with $A^2 = \frac34$.
The inflationary trajectory is now characterized by:
\begin{equation}
\langle \phi \rangle = 0, \quad T = \bar{T} \quad \text{(real modulus)},
\end{equation}
and we must now use the field redefinition in Eq.~(\ref{canT})
to normalize the inflaton kinetic term.

Using Eq.~(\ref{Wfphi}), the T-models are even more easily obtained by choosing
\beq
f\left(\frac{\phi}{M_P} \right) = \left(\frac{\phi}{M_P} \right)^{\frac{k}{2}} \left(3-\frac{\phi^2}{M_P^2}\right) \,,
\eeq
with $A^2 = \frac14 3^{-k/2}$. The corresponding function of $f(T)$ can be read off from Eq.(\ref{WTk}). 

When $\alpha \ne 1$, the choice of superpotential is somewhat more complicated. We we can use the same form for the superpotential given in Eq.~(\ref{Wfphi}) or Eq.~(\ref{simpT}) and obtain
\beq
V = 3^{3\alpha-2} A^2 \lambda M_P^4 \frac{|f(\phi/M_P)|^2}{\alpha \left(3-|\phi/M_P|^2\right)^{(3\alpha-1)}} \, ,
\eeq
or
\beq
V = 4 A^2 \lambda M_P^4 \frac{|T/M_P|^2 |f(T/M_P)|^2}{\alpha (T/M_P+{\bar T}/M_P)^{3\alpha-1}} \, ,
\eeq
respectively. The generalized E-model (\ref{eq:alphastarogeneral}) is then obtained with either
\beq
f(\phi/M_P) = \left(\frac{\phi/M_P}{\phi/M_P+\sqrt{3}}\right)^{\frac{k}{2}} \left(3-\frac{\phi^2}{M_P^2}\right)^{(\frac32 \alpha - \frac12)} \, ,
\eeq
or
\beq
f(T/M_P) = \left(1-\frac{M_P}{2T} \right)^\frac{k}{2} \left(\frac{T}{M_P} \right)^{(3 \alpha -3)/2} \, ,
\eeq
with $A^2 = 2^{k-2} 3^{(3-3\alpha)} \alpha$ or $A^2 = 3 \cdot 2^{(3\alpha-5)} \alpha$ and using the canonical field normalization 
\begin{equation}
    \label{eq:phicanonical1}
    \phi \; = \; \sqrt{3} M_P \tanh(\frac{\varphi}{\sqrt{6 \alpha} M_P}) \, ,
\end{equation} or 
\begin{equation}
\label{eq:tcanonicalfield}
T = \bar{T} = \frac{1}{2}e^{\sqrt{\frac{2}{3\alpha}} \frac{\varphi}{M_P}}\,,
\end{equation}
to obtain the canonical inflaton.

Similarly, the generalized T-model potential (\ref{eq:tmodelgeneral}) also follows from either (\ref{Wfphi}) or Eq.~(\ref{simpT}) with a different choice of superpotential functions
\beq
f(\phi/M_P) = \left(\frac{\phi}{M_P} \right)^{\frac{k}{2}} (3-\phi^2/M_P^2)^{(\frac32 \alpha - \frac12)} \, ,
\eeq
or
\beq
f(T/M_P) = \left( \frac{2T/M_P-1}{2T/M_P+1} \right)^{\frac{k}{2}} \left(\frac{T}{M_P} \right)^{(3 \alpha -3)/2} \, .
\eeq 
Then, using  $A^2 = \frac14 3^{(3-3\alpha-\frac{k}{2})} \alpha$ or $A^2 = 3 \cdot 2^{(3\alpha-5)} \alpha$ and Eq.~(\ref{eq:phicanonical1}) or Eq.~(\ref{eq:tcanonicalfield}) we recover the potential in Eq.~(\ref{eq:tmodelgeneral}).

Finally, we show how to construct Starobinsky-like and T-models with small deformations that lead to larger values of the spectral tilt $n_s$, which can more easily accommodate the ACT DR6 results.

In the case of the modified Starobinsky-like models, we use
\beq
\begin{aligned}
f(\phi/M_P) = \left(\frac{\phi}{M_P} \right)^{k/2} \left(3 -\frac{\phi^2}{M_P^2}\right)^{\frac{1}{2}(3\alpha - 1 - k)} \\
\times \left(\kappa \frac{\phi}{M_P} -\sqrt{3}\right)^{k/2} \, ,
\end{aligned}
\eeq
with $A^2 = 2^{k-2} 3^{3-3\alpha} \alpha$, or
\beq
f(T/M_P) = \frac{\left(-1+4 T^2/M_P^2 -\kappa(1-2 T/M_P)^2\right)^{\frac{k}{2}}}{ (T/M_P)^{\frac12(3+k-3\alpha)}} \, ,
\eeq
with $A^2 = 3 \cdot 2^{(3\alpha-5-2k)} \alpha$.
Evaluating the scalar potential along the inflationary trajectory in either case leads to Eq.~(\ref{eq:modstaro}). 
For $\kappa = 1$, we recover Eq.~(\ref{eq:alphastarogeneral}).

For the modified T-models, we use 
\beq
f(\phi/M_P) = \left(3 - \phi^2/M_P^2\right)^{\frac{1}{2}(3\alpha - 3)} \left(\frac{\phi}{M_P}\right)^{\frac{k}{2}}
\left( 1 - \frac{\kappa}{3} \frac{\phi^{2}}{M_P^2} \right)  \, ,
\eeq
with $A^2 = \frac14 3^{5-3 \alpha - \frac{k}{2}} \alpha $, or
\beq
\begin{aligned}
f(T/M_P) = \left( \frac{2T/M_P-1}{2T/M_P+1} \right)^{\frac{k}{2}}\left( \frac{T}{M_P}\right)^{(3 \alpha -5)/2} \\
\times \left( (1+2T/M_P)^2-\kappa (1-2T/M_P)^2 \right) \, ,
\end{aligned}
\eeq
with $A^2 = 3 \cdot 2^{3\alpha -11} \alpha$ which, combined with Eq.~(\ref{eq:phicanonical1}) or Eq.~(\ref{eq:tcanonicalfield}), leads to Eq.~(\ref{eq:modtmodel}) and 
reduces to Eq.~(\ref{eq:tmodelgeneral}) when $\kappa = 1$.

\subsection{Analytical Approximations for Inflationary Observables}
\label{appB}

While the results presented in the main text are obtained through numerical integration, we provide here analytical approximations that offer insight into the parameter dependencies and serve as useful benchmarks for numerical calculations.

Solving the condition (\ref{epsV}) for the $\alpha$-attractor potentials yields the following field values at the end of inflation. For E-models:
\begin{align}
\label{eq:phiendE}
\frac{\varphi_{\rm{end}}}{M_P} & \; \simeq \; \sqrt{\frac{3\alpha}{2}} \ln\left[\frac{2(6\alpha + 3\sqrt{3\alpha}-2)}{12\alpha-1} \right] \,, & 
\end{align}

and for T-models:
\begin{align}
\label{eq:phiendT}
\frac{\varphi_{\rm{end}}}{M_P} & \; \simeq \;  \sqrt{\frac{3\alpha}{2}}  \ln\left[ \frac{4-6\sqrt{\alpha(5+4\alpha)}}{1-12\alpha}\right.  & \nonumber\\
& \qquad\qquad\qquad +\left.\sqrt{\frac{75}{5+68\alpha+16\sqrt{\alpha(5+4\alpha)}}}\right] \, . & 
\end{align}
For the canonical case $\alpha = 1$, these expressions reduce to:
\begin{align}
\text{$\alpha$-Starobinsky:} \quad \varphi_{\mathrm{end}}/M_P &= 0.63 \,, \\
\text{T-model:} \quad \varphi_{\mathrm{end}}/M_P &= 0.89 \,,
\end{align}
recovering the values given in Eqs.~(\ref{alphastarophiend}) and~(\ref{tmodelphiend}).

The field value $\varphi_*$ when the pivot scale exits the horizon is determined by integrating the number of $e$-folds from Eq.~(\ref{eq:efolds}). The analytical solutions for the $\alpha$-Starobinsky model are:
\begin{align}
\label{eq:phistarstaro} \notag
\frac{\varphi_{*}}{M_P} & \; \simeq \; \sqrt{\frac{3\alpha}{2}}
\left[1 + \frac{3\alpha}{4N_*-3\alpha} \right] \\
& \qquad \times
\ln\left(\frac{4N_*}{3\alpha} + e^{\sqrt{\frac{2}{3}} \frac{\varphi_{\rm{end}}}{M_P}} - \sqrt{\frac{2}{3}} \frac{\varphi_{\rm{end}}}{M_P} \right)  \,, 
\end{align}
and for the T-model:
\begin{align}
\label{eq:phistarT}
\frac{\varphi_{*}}{M_P} & \; \simeq \;  \sqrt{\frac{3\alpha}{2}}  \cosh^{-1} \left[\frac{4N_*}{3\alpha} + \cosh\left(\sqrt{\frac{2}{3\alpha}} \frac{\varphi_{\rm{end}}}{M_P} \right) \right] \, ,
\end{align}
which reduce to Eqs.~(\ref{alphastarophistar}) and (\ref{tmodelphistar}) when $\alpha = 1$.
We have compared these analytical approximations with exact numerical results across the parameter ranges relevant for CMB constraints. The relative errors are summarized in Table~\ref{tab:accuracy}.

\begin{table}[h]
\centering
\caption{Maximum relative errors in the analytical approximations for $40 < N_* < 60$.}
\label{tab:accuracy}
\vspace{3mm}
\begin{tabular}{l|c|c|c}
\hline\hline
Model & $\alpha = 0.1$ & $\alpha = 1$ & $\alpha = 10$ \\
\hline
\multicolumn{4}{c}{$\varphi_{\mathrm{end}}/M_P$} \\
\hline
$\alpha$-Starobinsky & 2\% & 2\% & 4\% \\
T-model & 3\% & 5\% & 5\% \\
\hline
\multicolumn{4}{c}{$\varphi_*/M_P$} \\
\hline
$\alpha$-Starobinsky & 0.3\% & 0.3\% & 3\% \\
T-model & 0.4\% & 0.5\% & 0.7\% \\
\hline\hline
\end{tabular}
\end{table}

The approximations for $\varphi_*$ are particularly accurate, with errors below 1\% for $\alpha \lesssim 1$. The larger errors for $\varphi_{\mathrm{end}}$ at large $\alpha$ values reflect the breakdown of the slow-roll approximation near the end of inflation. Nevertheless, these analytical expressions provide reliable estimates for cosmological observables and serve as efficient starting points for numerical calculations.

The post-inflationary evolution significantly affects the relationship between the number of $e$-folds $N_*$ and the inflationary parameters. 
For $k=2$, simplification of Eq.~(\ref{eq:nstarreh}) for $N_*$ as a function of the reheating temperature is possible using the relation~\cite{Ellis:2021kad}
\beq
\frac{1-3w_{\mathrm{int}}}{12(1+w_{\mathrm{int}})}\ln\left(\frac{\rho_{\mathrm{rad}}}{\rho_{\mathrm{end}}}\right)\;\simeq\;\frac{1}{6}\ln\left(\frac{\Gamma_{\varphi}}{H_{\rm end}}\right)\,.
\eeq
At the pivot scale $k_* = 0.05$ Mpc$^{-1}$ we can therefore write
\begin{align} \notag
N_* \;\simeq\; &61.41 + \frac{1}{6}\ln\lambda + \frac{1}{2}\ln \left( \frac{V_*}{\lambda M_P^4}\right)  \\ \label{eq:Nstark2}
& - \frac{1}{3} \ln \left( \frac{V_{\rm end}}{\lambda M_P^4}\right) + \frac{1}{3}\ln\left(\frac{T_{\rm RH}}{M_P}\right) \,,
\end{align}
where the dependence on $\alpha$ can be obtained by substitution of (\ref{eq:phiendE}) and (\ref{eq:phistarstaro}) into the $\alpha$-Starobinsky potential (\ref{eq:emodel}), or by substitution of (\ref{eq:phiendT}) and (\ref{eq:phistarT}) into the T-model potential (\ref{eq:tmodel}). The dependence on the coupling $\lambda$ is explicitly separated, and requires the substitution (\ref{eq:infnorm}). For $\alpha=1$ this simplifies to
\beq
N_* \;\simeq\; 59.55 - \frac{1}{3}\ln N_* + \frac{1}{3}\ln\left(\frac{T_{\rm RH}}{M_P}\right) \,,
\eeq
for $\alpha$-Starobinsky models, and for T-models
\beq
N_* \;\simeq\; 59.67 - \frac{1}{3}\ln N_* + \frac{1}{3}\ln\left(\frac{T_{\rm RH}}{M_P}\right) \,.
\eeq
Comparison with full numerical calculations shows that these analytical approximations achieve remarkable accuracy. We find a maximal error of 0.2\% for $\alpha$-Starobinsky models and a maximal error of 0.1\% for T-models across the parameter range shown in Fig.~\ref{fig:NvsTfull}. These analytical results provide important insight into how reheating dynamics affects inflationary predictions, demonstrating that uncertainties in the reheating temperature translate into relatively modest shifts in $N_*$ due to the logarithmic dependence. For comprehensive analyses of reheating in $\alpha$-attractor models, see~\cite{Drewes:2017fmn,German}.

For $k>2$ Eq.~(\ref{epsV}) can be solved to obtain approximate expressions for the inflaton value at the end of inflation. For the E-models this yields 
\beq\label{eq:phiendEk}
\frac{\varphi_{\rm end}}{M_P}\;\simeq\; \sqrt{\frac{3\alpha}{2}} \ln\left[\frac{2(6\alpha + \sqrt{3\alpha(4k^2-4k+1)}-k)}{12\alpha-1} \right]\,.
\eeq
while for T-models
\begin{align}\notag
\frac{\varphi_{\rm end}}{M_P}\;&\simeq\; \sqrt{\frac{3\alpha}{2}}  \ln\left[ \frac {2k(1-2\sqrt{3\alpha})} {1 - 12\alpha}   + \right.\\ \label{eq:phiendTk}
&\ \left. \sqrt {\frac {\left (1 + 12\alpha - 4\sqrt{3\alpha} \right)\left (12\alpha + 4 k^2 - 1 \right)} {(12\alpha - 
       1)^2}} \right]\,.
\end{align}
Similarly, the corresponding expressions for the field value at the horizon exit of the pivot scale are given by
\begin{align}
\notag
\frac{\varphi_{*}}{M_P} & \; \simeq \; \sqrt{\frac{3\alpha}{2}}
\left[1 + \frac{3\alpha}{2kN_*-3\alpha} \right] \\
& \qquad \times
\ln\left(\frac{2kN_*}{3\alpha} + e^{\sqrt{\frac{2}{3}} \frac{\varphi_{\rm{end}}}{M_P}} - \sqrt{\frac{2}{3}} \frac{\varphi_{\rm{end}}}{M_P} \right)  \,, 
\end{align}
for the E-models, and 
\beq
\frac{\varphi_{*}}{M_P} \; \simeq \;  \sqrt{\frac{3\alpha}{2}}  \cosh^{-1} \left[\frac{2kN_*}{3\alpha} + \cosh\left(\sqrt{\frac{2}{3\alpha}} \frac{\varphi_{\rm{end}}}{M_P} \right) \right] \, ,
\label{eq:phistarTk}
\eeq
for the T-models. With these expressions at hand, the equivalent of Eq.~(\ref{eq:Nstark2}) takes the form
\begin{align}\notag
N_* \;\simeq\; &61.49 + \ln\left[ \left(\frac{3}{2}\right)^{\frac{k+2}{6k}} \left(\frac{\pi^2}{30}\right)^{\frac{k+4}{12k}} \right] + \frac{k-1}{3k}\ln\lambda\\ \notag
&+ \frac{1}{2} \ln \left( \frac{V_*}{\lambda M_P^4}\right) - \frac{k+2}{6k} \ln \left( \frac{V_{\rm end}}{\lambda M_P^4}\right)\\ \label{eq:Nstarka}
& - \frac{k-4}{3k}\ln\left(\frac{\max[T_{\rm RH},T_{\rm frag}]}{M_P}\right) - \frac{k-2}{6k}\ln g_{\rm RH}\,,
\end{align}
where $T_{\rm frag}$ denotes the temperature of the radiation at the moment of inflaton fragmentation originated from the self-interaction of the inflaton (see the corresponding discussion in Section~\ref{sec:phik}). We note that for $k=4$ and $\alpha=1$ this gives $N_*\simeq 55.7$ (E-model) and $N_*\simeq 55.8$ (T-model) for the MSSM degrees of freedom, independent of the details of reheating.

\subsection{Numerical Computation of Inflationary Observables}
\label{appC}
\setcounter{equation}{0}

While the slow-roll approximations presented in Section~\ref{sec:infdynamics} provide valuable analytical insights, precise comparison with CMB data requires numerical integration of the perturbation equations. This is particularly important for models near the boundary of observational constraints, where percent-level accuracy in $n_s$ can determine viability.

We compute the primordial scalar power spectrum by solving the equation of motion for gauge-invariant curvature perturbations. In the uniform-density gauge, the Mukhanov-Sasaki variable is defined as:
\begin{equation}
Q_k = \delta\varphi_k + \frac{\dot{\varphi}}{H}\Psi_k,
\end{equation}
where $\delta\varphi_k$ and $\Psi_k$ are the Fourier modes of the inflaton and metric perturbations, respectively. The evolution equation for $Q_k$ follows from the perturbed Einstein equations~\cite{Lalak:2007vi,egno3}:
\begin{equation}
\label{eq:MSeq}
\ddot{Q}_k + 3H\dot{Q}_k + \left[\frac{k^2}{a^2} + m_{\text{eff}}^2\right]Q_k = 0 \, ,
\end{equation}
where the effective mass term is:
\begin{equation}
m_{\text{eff}}^2 = V_{\varphi\varphi} - \frac{\dot{\varphi}^4}{2 H^2} + \frac{2\dot{\varphi}V_{\varphi}}{ H} + 3\dot{\varphi}^2 \,.
\end{equation}
We impose Bunch-Davies initial conditions in the sub-horizon limit:
\begin{equation}
Q_k = \frac{1}{a\sqrt{2k}}e^{-ik\tau} \quad \text{for} \quad k \gg aH \, ,
\end{equation}
where $\tau = \int dt/a$ is the conformal time. The comoving curvature perturbation is related to the Mukhanov-Sasaki variable by
\begin{equation}
\mathcal{R}_k = \frac{H}{\dot{\varphi}}Q_k \,.
\end{equation}
The dimensionless power spectrum, defined through
\begin{equation}
\langle\mathcal{R}_k\mathcal{R}_{k'}^*\rangle = \frac{2\pi^2}{k^3}\mathcal{P}_{\mathcal{R}}(k)\delta^{(3)}(\mathbf{k} - \mathbf{k'}),
\end{equation}
is evaluated numerically at horizon crossing ($k = aH$) for each mode. The spectral index is then computed from its definition at the end of inflation:
\begin{align}
\label{eq:nsdef}
n_s - 1 &= \frac{d\ln\mathcal{P}_{\mathcal{R}}}{d\ln k}\bigg|_{k=k_*}.
\end{align}

Analogously, we solve numerically for the tensor perturbations, using the standard transverse, traceless perturbation. The equation of motion satisfied by the tensor mode functions is given by
\beq\label{eq:heq}
\ddot{h}_{k,\gamma} + 3H\dot{h}_{k,\gamma} + \frac{k^2}{a^2} h_{k,\gamma} \;=\; 0\,.
\eeq
where $\gamma=+,\times$ denotes the two polarization states. The corresponding power spectrum is defined as
\beq
\sum_{\gamma=+,\times} \langle h_{k,\gamma}h^{\dagger}_{k',\gamma}\rangle \;=\; \frac{2\pi^2}{k^3}\mathcal{P}_{\mathcal{T}}(k)\delta^{(3)}(\mathbf{k} - \mathbf{k'}),
\eeq
with the tensor-to-scalar ratio computed then as follows,
\beq
\label{eq:rdef}
r = \frac{\mathcal{P}_{\mathcal{T}}}{\mathcal{P}_{\mathcal{R}}}\bigg|_{k=k_*}.
\eeq

\begin{figure}[ht!]
\centering 
\includegraphics[width=\columnwidth]{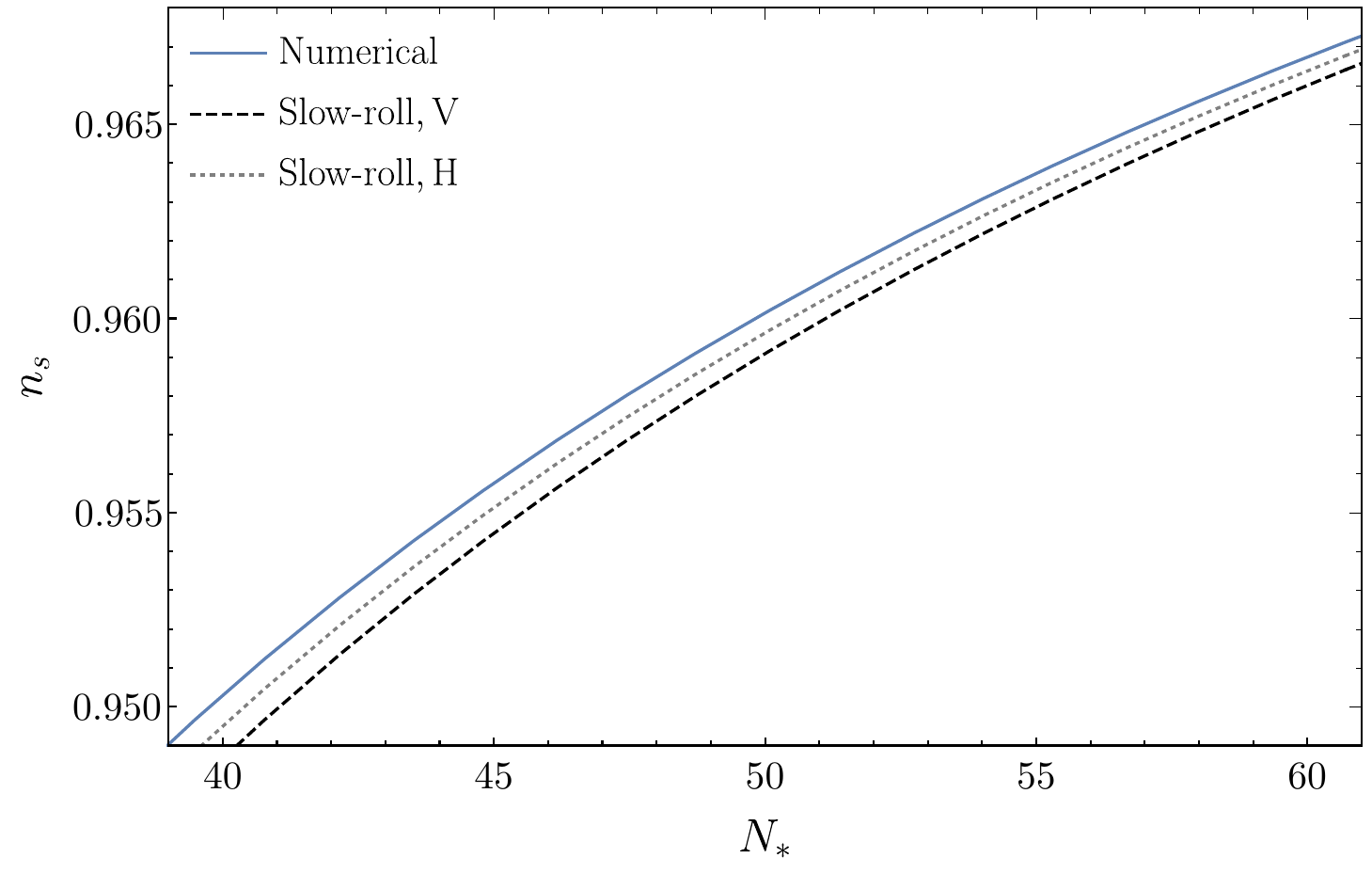}
\caption{The spectral tilt, $n_s$, as a function of the number of $e$-folds $N_*$ for the $\alpha$-Starobinsky model with $\alpha = 1$. The solid blue line shows the exact numerical solution of the Mukhanov-Sasaki equation~(\ref{eq:MSeq}). The dotted gray line represents the slow-roll approximation using Hubble flow parameters~(\ref{eq:epsetaH}), while the dashed black line uses potential slow-roll parameters~(\ref{eq:epseta}). The shaded region indicates the 68\% CL constraint from \textit{Planck} 2018 data.}
\label{fig:nsN}
\end{figure}

Our numerical results reveal appreciable corrections to the slow-roll predictions. Fig.~\ref{fig:nsN} illustrates the relationship between $n_s$ and $N_*$ computed using three methods:

\begin{enumerate}
\item \textbf{Full numerical integration} (solid line): Exact solution of Eq.~(\ref{eq:MSeq});
\item \textbf{Potential slow-roll} (dashed line): Using parameters from Eq.~(\ref{eq:epseta});
\item \textbf{Hubble slow-roll} (dotted line): Using the Hubble flow parameters:
\begin{align}
\label{eq:epsetaH}
\varepsilon_H &= -\frac{\dot{H}}{H^2} = \frac{\dot{\varphi}^2}{2M_P^2 H^2}, \\
\eta_H &= 2\varepsilon_H - \frac{\dot{\varepsilon}_H}{2\varepsilon_H H}.
\end{align}
\end{enumerate}

The potential slow-roll approximation exhibits systematic errors of $\Delta N_* \gtrsim 1$ when matching to a given value of $n_s$. This discrepancy arises from violations of slow-roll near the end of inflation, higher-order gradient corrections in the perturbation equations, and the evolution of slow-roll parameters during horizon crossing. The Hubble slow-roll approximation provides improved accuracy as it better captures the instantaneous dynamics. For precision cosmology, however, full numerical integration remains essential, particularly when confronting models with tight observational constraints such as those from ACT DR6.

The accuracy of different approximation schemes becomes apparent when comparing predictions for the spectral tilt as a function of $e$-fold number. Fig.~\ref{fig:nsN} demonstrates these differences for the canonical Starobinsky model ($\alpha = 1$).

\end{document}